\documentclass[onecolumn, showpacs,preprintnumbers,amsmath,amssymb,11pt,a4paper]{revtex4-1}

\usepackage{geometry}

\usepackage{hyperref}

\usepackage{physics}
\usepackage{float}
\usepackage{verbatim}
\usepackage{ wasysym }
\usepackage{graphicx}% Include figure files
\usepackage{dcolumn}% Align table columns on decimal point
\usepackage{bm}% bold math
\newcommand{\be}{\begin{equation}}
\newcommand{\ee}{\end{equation}}
\newcommand{\bd}{\begin{displaymath}}
\newcommand{\ed}{\end{displaymath}}
\newcommand{\BE}{\begin{eqnarray}}
\newcommand{\EE}{\end{eqnarray}}

\newcommand{\boldpsi}{{\mbox{\boldmath $\psi$}}}

\newcommand{\avg}[1]{\left\langle{#1}\right\rangle}

\usepackage[dvipsnames]{xcolor}
\usepackage{framed,color}
\definecolor{shadecolor}{rgb}{0.85,0.80,0.80}
\definecolor{myorange}{RGB}{253, 184, 99}
\definecolor{mypurple}{RGB}{178, 171, 210}

\usepackage{physics}
\begin{document}

\preprint{}
\title{Ecological communities from random generalised Lotka-Volterra dynamics with non-linear feedback}
% Force line breaks with \\

\author{Laura Sidhom}
\email{laura.sidhom@postgrad.manchester.ac.uk}

\affiliation{Theoretical Physics, Department of Physics and Astronomy, School of Natural Sciences, The University of Manchester, Manchester M13 9PL, United Kingdom}

\author{Tobias Galla}
\email{tobias.galla@manchester.ac.uk}

\affiliation{Theoretical Physics, Department of Physics and Astronomy, School of Natural Sciences, The University of Manchester, Manchester M13 9PL, United Kingdom}

\affiliation{Instituto de F\'isica Interdisciplinar y Sistemas Complejos, IFISC (CSIC-UIB),
Campus Universitat Illes Balears, E-07122 Palma de Mallorca, Spain}

\date{\today}% It is always \today, today,
             %  but any date may be explicitly specified

\begin{abstract}
We investigate the outcome of generalised Lotka-Volterra dynamics of ecological communities with random interaction coefficients and non-linear feedback. We show in simulations that the saturation of non-linear feedback stabilises the dynamics. This is confirmed in an analytical generating-functional approach to generalised Lotka-Volterra equations with piecewise linear saturating response. For such systems we are able to derive self-consistent relations governing the stable fixed-point phase, and to carry out a linear stability analysis to predict the onset of unstable behaviour. We investigate in detail the combined effects of the mean, variance and co-variance of the random interaction coefficients, and the saturation value of the non-linear response. We find that stability and diversity increases with the introduction of non-linear feedback, where decreasing the saturation value has a similar effect to decreasing the co-variance.
We also find co-operation to no longer have a detrimental effect on stability with non-linear feedback, and the order parameters mean abundance and diversity to be less dependent on the symmetry of interactions with stronger saturation.
\end{abstract}

\maketitle
%%%%%%%%%%%%%%%%%%%%%%%%%%%%%%%%%%%%%%%%%%%%%%%%%%%%%%%%%%%%%%%%%%%%%%%%%%%%%%%%
%%%%%%%%%%%%%%%%%%%%%%%%%%%%%%%%%%%%%%%%%%%%%%%%%%%%%%%%%%%%%%%%%%%%%%%%%%%%%%%%

\section{Introduction}

 The discussion whether large ecosystems can maintain stability and diversity or not has a long tradition \cite{elton,macarthur,gardner}. While models with random interaction parameters were introduced more than 45 years ago by May \cite{may1,may2}, they continue to play an important role in this diversity-stability debate. Models with random coupling coefficients are used not only for the modelling of large-scale ecosystems, but also to describe interactions in the human microbiome. For example, recent studies have examined how different types of interactions between microbe species, and between the human host and the microbes can affect stability \cite{kat,leash}.
  
 Several approaches have been taken to study the stability of ecological communities with random interaction matrices. One is concerned with assemblies with a fixed given size, $S$, and assumes that their interactions are set by a random matrix. More precisely, these models focus on the study of the eigenvalues of a putative $S\times S$ Jacobian matrix, which is assumed to be random. This line of approach has been taken for example in \cite{may1,may2} and in \cite{allesina,allesina2012,allesina2015,tang, grilli,gibbs}. Knowledge and ideas from statistical physics contribute to these studies, exploiting technology developed for random-matrix problems for example in nuclear physics, the theory of disordered systems or condensed matter physics \cite{mehta}.

A second approach focuses on {\em dynamical} models of species interaction. These are often based on coupled differential equations, governing the abundances of species and their interactions. Typical examples are generalised Lotka-Volterra equations, or closely related, replicator dynamics of evolutionary game theory. These involve a definition of reproductive growth rate, which in turn requires a notion of species-to-species interaction. One assumes for example pairwise interaction with fixed random coefficients. Several dynamical assembly approaches can then be taken, such as the step-wise assembly process. This has been studied in simulations \cite{post}, and also analytically \cite{wiley}. We use a separate approach, initiated originally in \cite{opper}. We start from a pool of species, who interact through a generalised Lotka-Volterra dynamics. In the course of these dynamics some species may become extinct, and we are interested in the community of remaining species. Given the fixed interaction coefficients, such models are dynamical problems with quenched disorder in the language of the theory of disordered systems (for a general reference see \cite{mpv}). 

Tools from equilibrium and non-equilibrium statistical physics can then be used to make analytical progress, usually relying on the assumption that the number of species in the system is large; formally the thermodynamic limit of an infinite number of species is taken. Different types of behaviour can then be found, both in simulations and from analytical approaches. For example dynamical systems of this type can approach stable fixed points, and in suitable parameter regimes these fixed points are found to be unique, i.e., independent of the initial conditions. For other parameters multiple fixed points or equilibria can be found. Their number and statistics can be characterised for example using Gardner-type calculations \cite{berg1,berg2}.

Static approaches to ecosystems are based on the celebrated replica method \cite{diederich,biscari,fontanari}; this assumes the existence of a Liapunov function and typically requires a symmetric interaction matrix.
A separate approach uses the cavity method or the Thouless-Anderson-Palmer (TAP)  equations, again originally developed in the context of spin glasses (see e.g. \cite{mpv}). The focus here is not on the actual dynamics of the ecosystem, but on the statistics of fixed points, and on their stability \cite{bunin2016,bunin2017,biroli}. The advantage relative to the replica approach is that symmetry of the interaction matrix is not required. 

Finally, dynamic generating functionals (path integrals) have been used to study large ecosystems with disordered couplings. These techniques were originally developed for spin systems \cite{dominicis,msr,sommers} (for more recent reviews see \cite{coolendyn,coolenbook1,sollich}) and then first applied to replicator dynamics by Opper and Diederich \cite{opper}, and by Berg and Weigt \cite{berg2}. This was then further developed in \cite{gallaasym,gallahebb,yoshino1,yoshino2,gallaepl}. The path-integral approach is dynamic, in principle, and results in an effective process for a representative species. This dynamic mean-field theory describes the time evolution of a typical degree of freedom after the average over the quenched disorder has been carried out. In most cases this effective process involves a non-Markovian retarded interaction kernel and coloured noise. This makes analytical solution difficult, in particular in transient regimes when dynamic order parameters are time dependent. 

The effective dynamics resulting from the generating-functional analysis can be studied numerically using the method by Eissfeller and Opper \cite{eissfeller} to simulate sample paths of the effective dynamics, see also \cite{felixroy} for further recent developments of methods to evaluate dynamical mean field theory. Analytical solutions are feasible when model parameters are such that the system converges to a unique stable fixed point, independent of the initial condition. It is then possible to derive self-consistent relations for the statistics of these fixed points and macroscopic order parameters. In the context of ecological communities these order parameters represent, among others, the fraction of species which survive the in the long run, the diversity of these species (e.g., the species abundance distribution), and the mean abundance at the fixed point. 

The theory also self-consistently predicts its own instability, i.e. from the fixed-point solutions one can identify combinations of parameters at which dynamical instabilities set in. A number of different concepts of stability are used in the ecology literature \cite{unveiling, babel}. In this paper we are mainly interested in identifying the parameter regime in which the dynamics is globally stable, i.e., it reaches one unique fixed point, irrespective of the initial condition. As we will briefly discuss later the type of instability we identify using the generating-functional approach can be related to so-called `structural stability' in ecology \cite{wiley, structural}.

Outside the globally stable regime one finds phases in which the dynamics converge to stable fixed points, but where different fixed points are reached for different starting points of the dynamics. We also find phases with persistent dynamics, such as limit cycles, heteroclinic cycles and chaos. For generalised Lotka-Volterra equations with random coupling matrices, finally, phases with unbounded growth can be identified \cite{bunin2016,bunin2017,biroli,gallaepl2018}. Similar behaviour has also been seen in the context of replicator equations \cite{gallaasym}; we note that unbounded growth is not possible for replicator equations.

Most existing generating-functional studies of replicator or generalised Lotka-Volterra models focus on cases in which the resulting effective process takes a simple form, resulting in linear fixed-point relations. These are typically models in which the (relative) growth of the abundance of one species depends linearly on the abundances of the other species. Examples can be found in \cite{opper,gallaepl2018,gallaasym}. One notable exception are so-called Sato-Crutchfield dynamics in the context of game learning \cite{sato, gallafarmer}. This describes situations in which players update their mixed strategies in response to moves by their opponents, and payoffs received. For the learning models in \cite{sato, gallafarmer} mixed strategies evolve in a way similar to species abundances in a population, but the fixed-point relations contain logarithmic terms.

In this paper we focus on an example of non-linear feedback. This is inspired by the idea that the influence on growth from interactions with other species may not have an unlimited effect, but instead saturate. This type of non-linear feedback is often modelled using Hill functions \cite{hill}, similar to Holling type-II functional response \cite{holling} in ecology.
The aim of our work is to investigate how this type of non-linear feedback affects the outcome of evolution of ecosystems with random interactions. Specifically we focus on the effects of the non-linear feedback on the phase of global stability. We show that the saturation of feedback increases the region with a unique stable fixed point. This stabilizes communities, and leads to more diversity than in the absence of saturation.

The remainder of this paper is organized as follows. In Sec.~\ref{sec:def} we define the general class of models we will be looking at, and we introduce the main control parameters. We then present results from numerical simulations of random generalised Lotka-Volterra communities with non-linear feedback in Sec.~\ref{sec:holling}; in particular we report the different types of behaviours seen, and how the main model parameters influence this behaviour. In Sec.~\ref{sec:gf} we then develop the generating functional for the model with general feedback function, report the resulting effective species process, and the self-consistent equations characterising the regime of unique stable fixed points. To make further analytical progress we then focus on the case of piecewise linear saturating functional feedback, and carry out a linear stability analysis. The predictions from the theory are tested in Sec.~\ref{sec:pg}, where we report detailed phase diagrams obtained from the path-integral analysis and from simulations. We then compare our results from the piecewise linear feedback function to results from simulations of the model with non-linear feedback in Sec.~\ref{compare}. In Sec.~\ref{sec:dependence} we discuss the role of the different ecological parameters, and in particular how saturation in the non-linear feedback affects the stability of the ecosystem. We summarise our findings in Sec.~\ref{sec:concl}.

\section{Model definitions}\label{sec:def}
We consider a pool of $N$ species, which we label by $i=1,\dots, N$. We write $x_i(t)$ for the abundance of species $i$ in the ecosystem at time $t$. The dynamics proceed in continuous time, governed by the generalised Lotka-Volterra equations
\be
\dot x_i(t) = r_ix_i(t)\left[K_i-x_i(t) + g\left(\sum_j\alpha_{ij}x_j(t)\right)\right].\label{eq:lv_general}
\ee
The quantity $r_i$ denotes the growth rate of species $i$, and $K_i$ is the carrying capacity for the species. In absence of interactions ($\alpha_{ij}=0$ for all $i,j$), the abundance $x_i$ can at most take value $x_i=K_i$. In the following we will set $r_i=1$ and $K_i=1$ for all species, following \cite{gallaepl2018,bunin2018}. The coefficients $\alpha_{ij}$ describe the interactions between the different species. In the context of random generalised Lotka-Volterra dynamics these are quenched disordered random variables; that is to say, they are chosen from a specified distribution at the beginning, but then remain fixed as the dynamics unfold. We will define the statistics of the $\alpha_{ij}$ below.

The function $g(\cdot)$ describes the  `feedback', i.e. how the growth of any one species is affected by the interaction with the other species. We note that Eq.~(\ref{eq:lv_general}) is restricted to functional forms of the type $g\left(\sum_j\alpha_{ij}x_j\right)$. In principle, more general forms can be devised, but these are harder to analyse. As we will see below, the model defined in Eq.~(\ref{eq:lv_general}) produces intricate behaviour, and is sufficient to highlight some of the effects of non-linear feedback.

Random generalised Lotka-Volterra communities with linear functional feedback, $g(u)=u$, have for example been studied in \cite{bunin2016,bunin2017,biroli,gallaepl2018}. We here focus on non-linear functions $g$. We generally assume that $g$ is a non-decreasing function of its argument.

The coefficient $\alpha_{ij}$ denotes the reproductive benefit or detriment species $i$ receives when interacting with species $j$. We set $\alpha_{ii}=0$; self-interaction is already accounted for in Eq.~(\ref{eq:lv_general}). In our model we assume that the off-diagonal coefficients are drawn from a Gaussian distribution with the following statistics,
\BE\label{eq:mean_variance}
\overline{\alpha_{ij}}&=&\frac{\mu}{N},\nonumber \\
\overline{\alpha_{ij}^2}-\frac{\mu^2}{N^2}&=&\frac{\sigma^2}{N}.
\EE
In-line with literature on disordered systems \cite{mpv} we use an overbar to indicate the average over the quenched random variables $\{\alpha_{ij}\}$. Eqs.~(\ref{eq:mean_variance}) indicate that the mean of each matrix element is $\mu/N$, and their variance is $\sigma^2/N$. The scaling with $N$ is standard in the context of disordered systems, and is chosen to guarantee a meaningful thermodynamic limit $N\to\infty$, which we will eventually assume in the generating-functional analysis. 

The parameter $\mu$ controls the `baseline' interaction between any pair of species. Negative values of $\mu$ indicate a generally competitive environment; the presence of any species leads to negative feedback on the growth of the other species. Similarly, if $\mu$ takes positive values, species generally interact positively with each other, and the presence of one species tends to enhance the growth of all other species. One expects this to potentially lead to unlimited growth as in \cite{bunin2016,bunin2017}, at least in the absence of saturation effects in the non-linear feedback. We will refer to $\mu$ as the co-operation parameter. This is also known as the competition--mutualism parameter in ecology \cite{mutualism}.

The parameter $\sigma$ describes the degree of heterogeneity in the interaction of species, we will call it the `heterogeneity parameter'. We also allow for correlations between the interaction coefficients $\alpha_{ij}$ and $\alpha_{ji}$ for any pair $i\neq j$ of species. Specifically, we write these as follows
\be
\overline{\alpha_{ij}\alpha_{ji}}-\frac{\mu^2}{N^2}=\gamma\frac{\sigma^2}{N},
\ee
where the model parameter $\gamma$ can take values between $-1$ and $1$. The role of $\gamma$ can best be understood by considering the case $\mu=0$. In this case $\overline{\alpha_{ij}\alpha_{ji}}=\gamma\sigma^2/N$. For $\gamma=-1$ one then finds $\alpha_{ij}=-\alpha_{ij}$ with probability one, i.e. species form predator-prey pairs; one species in each pair benefits from the presence of the other, but that other species is adversely affected by the presence of the first. For $\gamma=0$ (and still assuming $\mu=0$) the interaction coefficients $\alpha_{ij}$ and $\alpha_{ji}$ are uncorrelated, i.e. half of all pairs of species will be of the predator-prey type, and the other half will either both benefit from each other, or each be suppressed by the presence of the other species. For $\gamma=1$ finally, there are no predator-prey pairs. Instead, $\alpha_{ij}=\alpha_{ji}$ with probability one, i.e. both species $i$ and $j$ profit from each other's presence, or the interaction is negative in both directions. If $\mu \neq 0$, the combination of $\alpha_{ij},\alpha_{ji}$ is drawn from a bivariate Gaussian distribution with non-zero mean, and the number of competitive, co-operative, and exploitative interactions can be obtained from the probabilities in the different quadrants of the $\alpha_{ij} - \alpha_{ji}$-plane. We will call $\gamma$ the symmetry parameter.

The main objective of our work is to investigate how the parameters $\mu$, $\sigma$ and $\gamma$ affect the outcome of the generalised Lotka-Volterra dynamics in the presence of non-linear feedback.

\section{Numerical results for non-linear feedback}\label{sec:holling}

We first focus on a model similar to Holling type-II functional response \cite{holling1,holling2}. This form of feedback was originally introduced to model the rate of growth of a predator while interacting with prey; it is natural that the benefit from additional prey will eventually saturate when prey numbers are large. We extend the idea of a saturating function to all types of inter-species interactions, and study the following feedback function,
 \be\label{eq:holling}
g(u) =g_H(u) \equiv \frac{2au}{a+2|u|}.
\ee
The subscript $H$ stands for Hill function. This function has a sigmoidal shape, and saturates to $g_H=a$ for $u\gg 1$ and to $g_H=-a$ for $u\ll -1$. We note that $g_H(u=\pm a/2)=\pm a/2$, i.e. the half-point of saturation is reached at $u=\pm a/2$. More general forms of the Hill function can be considered, but we here restrict the analysis to the form in Eq.~(\ref{eq:holling}), with one single parameter $a$.

In numerical simulations of the generalised Lotka-Volterra system with this type of feedback we broadly find three different dynamical outcomes: In some cases the system converges to a unique fixed point. That is to say, for a fixed draw of the interaction matrix elements $\{\alpha_{ij}\}$ the dynamics converge to one single fixed point, independent of what initial conditions are used for the $\{x_i\}$. In other cases we also find fixed points, but these are no longer unique. I.e., while runs are generally found to converge, the system has multiple marginally stable fixed-point attractors, and which one is eventually reached depends on the initial condition. The third type of outcome we find is one in which the dynamics never settles down and remains volatile until the end of the simulation. The general types of behaviour have been found previously in related systems, for example in random replicator systems, and in models of game learning \cite{gallaasym,gallafarmer,opper}.

For $a \to \infty$ we recover unrestricted linear feedback; the system can then display a fourth type of behaviour: unbounded growth \cite{gallaepl2018,bunin2016,bunin2017}. This is due to the lack of saturation. The absence of unlimited growth for finite values of the saturation parameter $a$ can directly be inferred from Eq.~(\ref{eq:lv_general}). For $r_i=K_i=1$ the relative growth rate of species $i$ is given by $\dot x_i/x_i=(1-x_i+g_i)$, where $g_i=g_H\left(\sum_j \alpha_{ij}x_j\right)$ at most takes value $g_i=a$. Thus the abundance of any species, $x_i$, is limited to at most $x_i=1+a$, as the growth rate for species $i$ then reduces to zero. We note the difference with random replicator dynamics \cite{opper,diederich,gallaasym}, in which the total abundance is constant in time ($N^{-1}\sum_i x_i=1$) by construction, but where none of the single variables $x_i$ is constrained to a fixed interval in the thermodynamic limit.

\begin{figure}[t!]
    \centering
    \includegraphics[width=\textwidth]{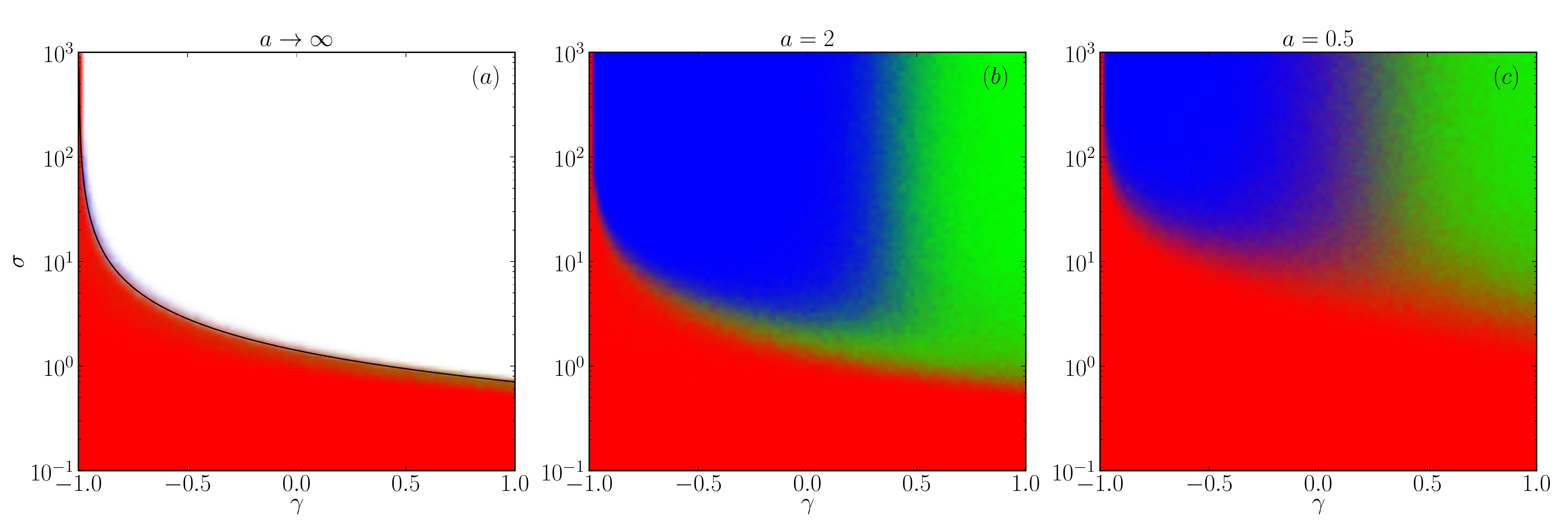}
    \caption{Phase diagram obtained from simulations of the generalised Lotka-Volterra system with $N=200$. Panel (a) shows the case $a\to\infty$ (i.e., linear feedback), panels (b) and (c) are for non-linear feedback, with $a=2$ in (b), and $a=0.5$ in (c). Simulations are for $\mu = 0$. The colours indicate the dominant outcome in each part of parameter space, with red (medium grey) representing a unique stable fixed point, green (light grey) multiple fixed points, blue (dark grey) indicating parameters for which the dynamics do not converge, and white indicating unbounded growth of species abundances. The solid line in panel (a) describes the onset of instability as derived from theory (see \cite{bunin2016,bunin2017,gallaepl2018}). }
    \label{holheat}
\end{figure}

We present numerical simulations for the generalised Lotka-Volterra system with non-linear feedback in Fig.~\ref{holheat}. The figures illustrates the behaviour of the system in the plane spanned by the symmetry parameter $\gamma$ and the heterogeneity parameter $\sigma$, for different values of the saturation parameter $a$. For each combination of these parameters we have carried out an ensemble of runs of the dynamics, and have recorded how frequently each dynamic outcome is observed. We describe how we distinguish between the three types of behaviour in the Supplementary Information. The frequencies with which the different outcomes are found are then converted into a colour (greyscale) code. Red colouring (medium grey) in the figure indicates parameters for which convergence to unique fixed points is found. In the green (light grey) areas of the phase diagram, we also observe predominantly convergence to fixed points, but the system has multiple such attractors, and which one is reached depends on the initial condition. In the blue (dark grey) areas of the graphs finally we find volatile behaviour, the trajectories generated by the generalised Lotka-Volterra system do not settle down by the end of our simulations. 

We broadly find the following phase behaviour. For sufficiently small heterogeneity $\sigma$ the system is stable and has a unique fixed point. This region of stability tends to be larger for low values of the symmetry parameter $\gamma$ than for higher values. The presence of predator-prey pairs (anti-correlation of the matrix elements) thus promotes the existence of a unique stable fixed point, in-line with previous observations for example in \cite{gallaepl2018,bunin2017}. If the heterogeneity $\sigma$ exceeds a critical value $\sigma_c$  the system either enters a phase with multiple stable fixed points, or it fails to converge. The latter tends to happen for lower values of $\gamma$, i.e. anti-correlated or moderately correlated interactions, the former for higher values of $\gamma$, when the interactions are increasingly correlated. The data in the figure shows that the saturation of the non-linear feedback stabilises the dynamics, with the globally stable region becoming larger for smaller values of the saturation parameter $a$ (i.e., for stronger non-linearity). For $a = \infty$ [panel (a) in Fig.~\ref{holheat}] we recover the unrestricted generalised Lotka-Volterra system with linear feedback, for which $\sigma_c^2$ is analytically known to be $\sigma_c^2=2/(1+\gamma)^2$ \cite{bunin2016,bunin2017,biroli,gallaepl2018}. This boundary is shown in Fig.~\ref{holheat}(a) as a solid line. For this particular choice of parameters the system shows unlimited growth in the unstable phase (indicated in Fig.~\ref{holheat}(a) by the absence of background shading).

We note that not all samples of the system show the same behaviour for any given set of parameters. For example some runs may converge, and others may fail to settle down before the end of the simulation. We believe that this is due to the finite number of species, finite integration time-step, and finite run-time required in simulations, and we would expect the boundaries to become more sharp in the asymptotic limit and for infinitely large systems. We also see trajectories which remain seemingly chaotic for a long time and then reach a fixed point only at very long times. The phase diagram in Fig.~\ref{holheat} shows the typical behaviour. Technically, we cannot exclude heteroclinic cycles based on our simulations, however we have not witnessed any such behaviour before the end-time of our numerical integration. In some regions of parameter space (most notably near the phase line), we also find that the number of simulations runs displaying a unique fixed point reduces when we increase the number of species $N$.

\section{Generating functional analysis}\label{sec:gf}
\subsection{Generating functional}
The generating-functional analysis proceeds along the lines of \cite{opper,gallaasym}. Starting from the dynamics in Eq.~(\ref{eq:lv_general}) one introduces the generating functional as
\be
Z[\boldpsi]=\avg{e^{i\sum_i\int dt~ x_i(t)\psi_i(t)}},
\ee
where $\avg{\dots}$ denotes an average over paths of the process; this includes averaging over potentially random initial conditions (we assume that the distributions of these are independent and identical across species). The field $\boldpsi$ is a source term. In essence $Z[\boldpsi]$ is the Fourier transform of the probability measure in the space of paths of the generalised Lotka-Volterra system. The generating functional is subsequently averaged over the quenched random coupling matrix, and the thermodynamic limit, $N\to\infty$ is taken. These steps are well-established, and the calculation is lengthy. We therefore only quote the final result here. Further intermediate steps are reported in the Supplementary Material.
\subsection{Effective representative species process}
The final outcome of the path-integral analysis, after the disorder has been averaged out and the thermodynamic limit has been taken, is an effective `mean-field' process for a single representative species. For the case of the generalised Lotka-Volterra dynamics in Eq.~(\ref{eq:lv_general}) (with $r_i=1, K_i=1$) the effective process is of the form
\BE
\dot x(t) &=& x(t)\bigg[1-x(t)+ g\left(\mu M(t)+\gamma\sigma^2 \int_0^t dt'~ G(t,t') x(t')+\eta(t)\right)\bigg],\label{eq:effproc}
\EE
where one has the following self-consistent relations  
\BE
M(t) &=& \left<x(t)\right>_*, \nonumber\\
\left<\eta(t)\eta(t')\right> &=& \sigma^2\left<x(t)x(t')\right>_*, \nonumber \\
G(t,t') &=& \left<\pdv{x(t)}{\eta(t')}\right>_*.\label{eq:selfc}
\EE
In these equations $\left<\cdots\right>_*$ denotes the average over realisations of the effective process, i.e., over the noise $\eta(t)$ and potentially random initial conditions. We will refer to $M(t)$, the correlation function $C(t,t')=\avg{x(t)x(t')}_*$, and the response function $G(t,t')$ as the macroscopic dynamical order parameters. In the context of ecology $M(t)$ is a measure of the average abundance in the system per species.

We note that the description in terms of the above effective process is also known as `dynamical mean field theory'. The process can alternatively be obtained using the cavity method, for models with linear feedback this is discussed in \cite{bunin2016, bunin2017, bunin2018, felixroy, biroli, barbier}.

\subsection{Fixed-point analysis}
We proceed to evaluate fixed points of the effective dynamics in Eq.~(\ref{eq:effproc}). The corresponding fixed-point relations are
\begin{align}
x^*\left[1 - x^*+g\left(\mu M^* + \gamma\sigma^2\chi x^* + \eta^*\right)\right] = 0,\label{eq:fp}
\end{align}
where we have used the superscript ${}^*$ to indicate quantities evaluated at the fixed point. In the fixed-point regime one has $G(t,t')=G(t-t')$ (time-translation invariance), and we have introduced the integrated response $\chi=\int_0^\infty d\tau~ G(\tau)$. The correlation function $C(t,t') = \left<x(t)x(t')\right>_*$ becomes independent of $t$ and $t'$ at the fixed point, and we write $q\equiv C(t,t')$. We can then replace $\eta^*$ by $\eta^*=\sigma\sqrt{q} z$, where $z$ is a static Gaussian random variable of mean zero and with variance one.
\smallskip

Eq.~(\ref{eq:fp}) always has the solution $x^*(z) = 0$ for all $z$. Potential other solutions fulfill the relation
\begin{align}
x^*=1+g\left(\mu M^* + \gamma\sigma^2\chi x^* + \sigma\sqrt{q}z\right) \label{eq:fp2}
\end{align}
Such solutions are only physically valid provided they are non-negative, as $x^*$ describes the abundance of an effective species. 

It is difficult to proceed analytically for general choices of the function $g$. In particular, if $g$ is non-linear Eq.~(\ref{eq:fp2}) would have to be solved numerically for $x^*(z)$ for a given value of $z$. We therefore consider a piecewise linear feedback function $g$. Specifically, we follow \cite{barbier} and approximate the non-linear feedback by
\begin{align}
g(u) = g_P(u)=
\begin{cases}
      a & u \geq a \\
      u & -a\leq u\leq a \\
      -a & u \leq -a.
 \end{cases}
 \end{align}
 The subscript $g_P$ refers to piecewise linear. This function is linear in $u$ in the interval $-a\leq u\leq a$, and is then `clipped'. Similar to the non-linear feedback the function saturates at $a$ for large $u$, and at $-a$ for large negative values of $u$. We also note that $g(a/2) = a/2$, i.e., the saturation half-point of the piecewise linear model is the same as for the non-linear feedback in Eq.~(\ref{eq:holling}). This structure allows us to proceed with the mathematical analysis, and at the same time it conserves some of the main features of the Holling type-II system as we will discuss further below. 

To find the solution $x^*(z)$ of Eq.~(\ref{eq:fp2}) we consider the three branches of $g_P(u)$. These are separated by threshold values $z_1$ and $z_2$ for the static noise variable $z$; we will evaluate these thresholds below. Specifically, we find:
\begin{itemize}
     \item[(i)] 
     For $z \geq z_2$, the argument of the function $g_P$ exceeds $a$ and hence $g_P(u) = a$; this gives the solution $x^*(z) = 1 + a$;
    \item[(ii)] For $z_1 \leq z \leq z_2$ one has $g_P(u) = u$, giving $x^*(z) = \frac{1 + \mu M^* +\sigma\sqrt{q}z}{1-\gamma\sigma^2\chi}$;
      \item[(iii)]
    For $z \leq z_1$ finally, the value $x^*(z)$ depends on the choice of the saturation parameter $a$ in the following way: if $a$ is smaller than the carrying capacity (i.e., $a\leq1$), the feedback saturates at $g_P=-a$, and we find $x^*(z)=1-a$.  If $a \geq 1$, the (effective) species dies out, $x^*(z) = 0$, before the feedback reaches saturation.
\end{itemize}

The threshold values $z_1$ and $z_2$ are found from the argument of the function $g$ in Eq.~(\ref{eq:fp2}),
\begin{align}\label{eq:zz}
z_1 &= \frac{(1-a)(1-\gamma\sigma^2\chi)\Theta(1-a) - (1 + \mu M^*)}{\sigma\sqrt{q}}, \nonumber \\
z_2 &= \frac{(1+a)(1-\gamma\sigma^2\chi) - (1 + \mu M^*)}{ \sigma\sqrt{q}}.
\end{align}
In the first expression $\Theta(\cdot)$ is the Heaviside step function, used here to differentiate between the cases $a \geq 1$ and $a \leq 1$. A more detailed discussion can be found in Sec.~S1~H of the Supplementary Information.

Putting the different cases together we find the following physical fixed-point value for a given combination of $z$ and $a$,
\be
 x^*(z) = \left\{\begin{array}{ccl}
    1+a &~&z \geq z_2, \\
    \frac{1 +\mu M^* + \sigma\sqrt{q}z}{1-\gamma\sigma^2\chi} & ~& z_1 \leq z \leq z_2, \\
    (1-a)\Theta(1-a)  & ~ & z \leq z_1.
\end{array}\right.
\label{eq:xcases}
\ee
 
We note that the abundance $x^*$ of the effective species is bounded from above by $1+a$. This indicates that, unlike in standard generalised Lotka-Volterra dynamics, abundances cannot diverge, and hence the average species population given by the order parameter $M^*$ also remains finite. The lower bound for the solutions of Eq.~(\ref{eq:fp2}) is zero for $a\geq 1$, and given by $1-a$ for $a \leq 1$.

It is important to recall that $x^*(z)=0$ is a solution of the fixed-point equation (\ref{eq:fp}) for all $z$. However, we find in linear stability analysis that this zero solution is an attractor only when $a\geq1$ and $z\leq z_1= -\frac{1 + \mu M^*}{\sigma\sqrt{q}}$, i.e. only when $x^*(z) = 0$ is the unique solution of the fixed point equation (\ref{eq:fp}). This is shown in Sec.~S1~I of the Supplementary Material. For $a<1$, the solution $x^*=0$ cannot be realised, and all species in the initial pool will have non-zero abundances in the phase of unique stable fixed points.

Using Eq.~(\ref{eq:xcases}) we can write Eqs.~(\ref{eq:selfc}) in the following form
\begin{subequations}\label{eq:opfp}
\begin{align}
M^* &= \int_{z_2}^\infty (a + 1) Dz + \int_{z_1}^{z_2} \frac{1 +\mu M^* + \sigma\sqrt{q}z}{1-\gamma\sigma^2\chi}Dz +\Theta(1-a) \int_{-\infty}^{z_1} (1-a) Dz, \label{eq:m} \\
q & = \int_{z_2}^\infty (a + 1)^2 Dz + \int_{z_1}^{z_2} \left(\frac{1 +\mu M^* + \sigma\sqrt{q}z}{1-\gamma\sigma^2\chi}\right)^2 Dz + \Theta(1-a)\int_{-\infty}^{z_1}(1-a)^2 Dz, \label{eq:q} \\
\chi & = \frac{1}{1-\gamma\sigma^2\chi} \int_{z_1}^{z_2}  Dz,\label{eq:chi}
\end{align}
\end{subequations}
where we have introduced the shorthand $Dz=\frac{dz}{\sqrt{2\pi}}e^{-z^2/2}$ for the Gaussian measure of $z$.
The quantity $M$ represents the mean species abundance, $q$ is the second moment of the species-abundance distribution, and $\chi$ is a `susceptibility', capturing how the fixed-point values of the species abundances shift in response to persistent perturbations.

Together with Eqs.~(\ref{eq:zz}) this is a self-consistent set of relations for the order parameters $q, \chi$ and $M^*$ in the regime of unique stable fixed points. Solutions of these equations can be obtained numerically as function of the model parameters $\mu, \sigma, \gamma$ and $a$. The method we use to solve this set of equations is described in the Section S3 of the Supplementary Material.

\subsection{Linear stability analysis}
We now carry out a linear stability analysis of the fixed points identified in the previous section. We first notice that fixed points of the form $x^*(z)=1+a$, $x^*(z)=(1-a)\Theta(1-a)$ are always locally stable. This is shown in Section S1~I in the Supplementary Material.  

We note that the function $g(\cdot)$ is the identity function in the vicinity of the remaining fixed points. We write $x(t)=x^*+y(t)$ and $\eta(t)=\sigma\sqrt{q}z+v(t)$, and following \cite{diederich} we add noise of zero mean and unit amplitude $\xi(t)$ in the effective process to study stability. 

Linearising the effective dynamics in Eq.~(\ref{eq:effproc}) in $y$ and $v$ we find
\be
\dot{y}(t) = x^*\left(-y(t)+\gamma\sigma^2\int_0^t dt' G(t-t) y(t')+v(t) +\xi(t)\right).\label{eq:help}
\ee
Carrying out a Fourier transform, this can be written as
\be
\frac{i\omega\tilde{y}(\omega)}{x^*} = \left(\gamma\sigma^2\tilde{G}(\omega) - 1\right)\tilde{y}(\omega) + \tilde{v}(\omega) + \tilde{\xi}(\omega).\label{eq:ft}
\ee
Following \cite{opper,gallaasym} we now focus on the long-time behaviour of perturbations, i.e., on the mode at $\omega=0$. This allows one to identify the transition to instability found in simulations, as described in more detail below. Broadly speaking, the behaviour of the $\omega=0$ mode is related to the long-term decay of perturbations. Discussions of stability conditions and the relevance of $\omega=0$-modes can also be found in \cite{sollich,felixroy}. Re-arranging Eq.~(\ref{eq:ft}) and taking averages we find
\be
\left<|\tilde{y}(0)|^2\right> = \phi \frac{\left<|\tilde{v}(0)|^2\right> + 1}{(1 - \gamma\sigma^2\chi)^2},
\ee
where the factor $\phi=\int_{z_1}^{z_2} Dz$ accounts for the fact that Eq.~(\ref{eq:help}) only applies to fixed points for which $g$ is not saturated, i.e. for values of $z$ with $z_1\leq z\leq z_2$. Finally, using the self-consistency relation $\avg{|\tilde y(0)|^2}=\sigma^2\avg{|\tilde v(0)|^2}$, we find
\begin{align}
\left<|\tilde{y}(0)|^2\right> = \frac{\phi}{(1-\gamma\sigma^2\chi)^2 - \phi\sigma^2}.
\end{align}
This quantity is finite (and non-negative) in the phase of stable fixed points, and becomes infinite when
\be\label{eq:instab}
\phi = \frac{(1-\gamma\sigma^2\chi)^2}{\sigma^2}.
\ee
This condition signals the onset of instability. Analogous conditions for Lotka-Volterra models with linear feedback have been related to stability conditions derived from random-matrix approaches, see for example \cite{opper,gallaepl2018,felixroy} for further discussion. This is related to what is known as `structural instability' in ecology, see \cite{structural, wiley}.

\section{Test against simulations and phase diagram}\label{sec:pg}

\subsection{Species abundance distributions}\label{sec:hist}

\begin{figure}[t!]
    \centering
    \includegraphics[width=0.85\textwidth]{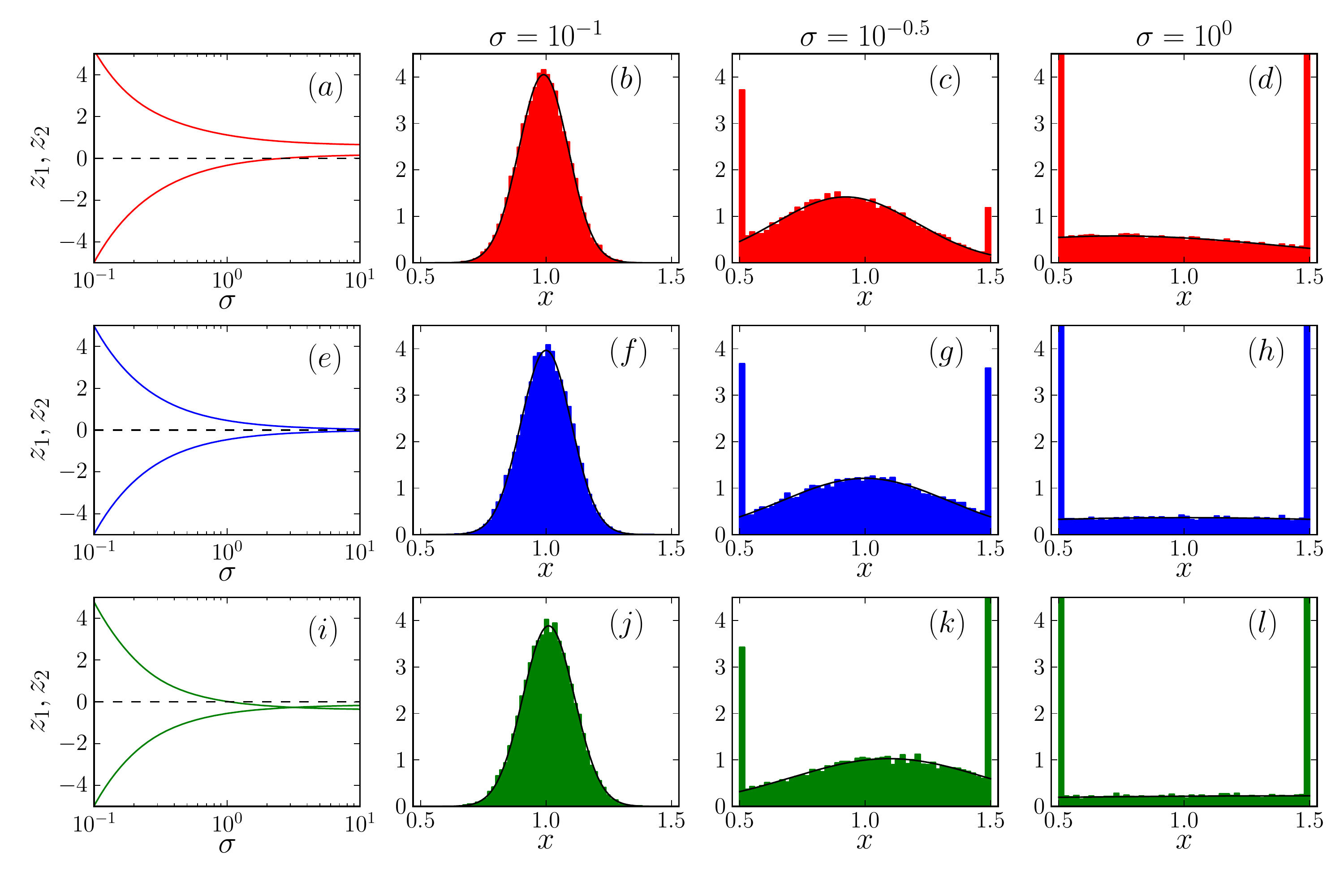}
    \caption{Species abundance distribution for the case $a=0.5$ and $\mu=0$ for different values of the heterogeneity parameter $\sigma$. The upper row [panels (a)-(d)] is for $\gamma=-1$, the middle row [(e)-(h)] for $\gamma=0$, and the lower row [(i)-(l)] for $\gamma=1$. On the left [(a),(e),(i)] we show $z_2$ (upper line) and $z_1$ (lower line), remaining panels show the species abundance distributions, for $\sigma=10^{-1}$ [(b),(f),(j)], $\sigma=10^{-0.5}$ [(c),(g), (k)] and $\sigma=1$ [(d),(h),(l)]. Solid lines are theoretical predictions for the species abundance distribution, shaded histograms are from simulations.}
    \label{hist5}
\end{figure}

We first discuss the resulting species abundance distributions in the regime of unique stable fixed points (i.e., the distribution of the $\{x_i\}$).
In Fig.~\ref{hist5} we show examples of species abundance distributions for different values of $\sigma$ and $\gamma$ for $a = 0.5$ and $\mu=0$. The shaded histograms in the figures are from simulations, solid black lines indicate the distributions of the unsaturated species from the theory. On the left of each of the figure [panels (a), (e), (i)] we show how $z_2$ (upper line) and $z_1$ (lower line) vary with increasing heterogeneity $\sigma$.

To interpret the histograms in Fig.~\ref{hist5} we note that the weight of each branch of the solution in Eq.~(\ref{eq:xcases}) is equal to the probability of a standard Gaussian random number $z$ to fall between or on either side of $z_1$ and $z_2$, respectively. For $\sigma = 0$, there is no species heterogeneity ($\alpha_{ij}=0$ for all $i,j$); all species abundances take the value $x^*=M^*$, where $M^* = 1/(1-\mu)$ from Eq.~(\ref{eq:m}). One finds $z_1\to -\infty$ and $z_2\to\infty$, and the non-linear feedback does not reach saturation.

For non-zero values of the interaction heterogeneity $\sigma$, $z_1$ and $z_2$ become finite; as a consequence there is a finite probability of $z$ falling outside the interval $[z_1,z_2]$, and hence the non-linear feedback saturates for a finite fraction of species. This results in a clipped Gaussian distribution for $x^*$; the fraction of species in the clipped regions increases with $\sigma$.
  
We find that both $z_1$ and $z_2$ are decreasing functions of $\gamma$, this is consistent with Eqs.~(\ref{eq:zz}), where the explicit factor of $\gamma$ dominates over the dependence of $M^*$, $q$ and $\chi$ on $\gamma$.
As a consequence the proportion of species `clipped off' at either side changes as $\gamma$ varies. We find a lower mean species abundance $M^*$, and more species at $1-a$ than at $1+a$ for $\gamma = -1$, and a higher mean abundance with more species at the upper bound $1+a$ for $\gamma = 1$.

We find similar results for $a=2$; these are shown in Fig.~S1 in the Supplementary Material. As the limiting values ($x^*=0$ and $x^*=1+a$) for the species abundances are further apart for this case, a higher value of $\sigma$ is required to spread the abundance distribution to these values. Therefore we find less saturation for $a = 2$ than for $a = 0.5$ at a fixed value of the heterogeneity parameter $\sigma$.

\subsection{Test of theoretical predictions for order parameters in the phase of unique stable fixed points}
The analytical theory results in predictions for the order parameters $q, \chi$ and $M^*$ as a function of the model parameters $a, \mu, \gamma$ and $\sigma$. These predictions are obtained as the solutions of the coupled equations (\ref{eq:m},\ref{eq:q},\ref{eq:chi}). They are valid in the parameter regime in which the generalised Lotka-Volterra system converges to a unique stable fixed point, independent of initial conditions.

\begin{figure}[t!]
    \centering
    \includegraphics[width=0.85\textwidth]{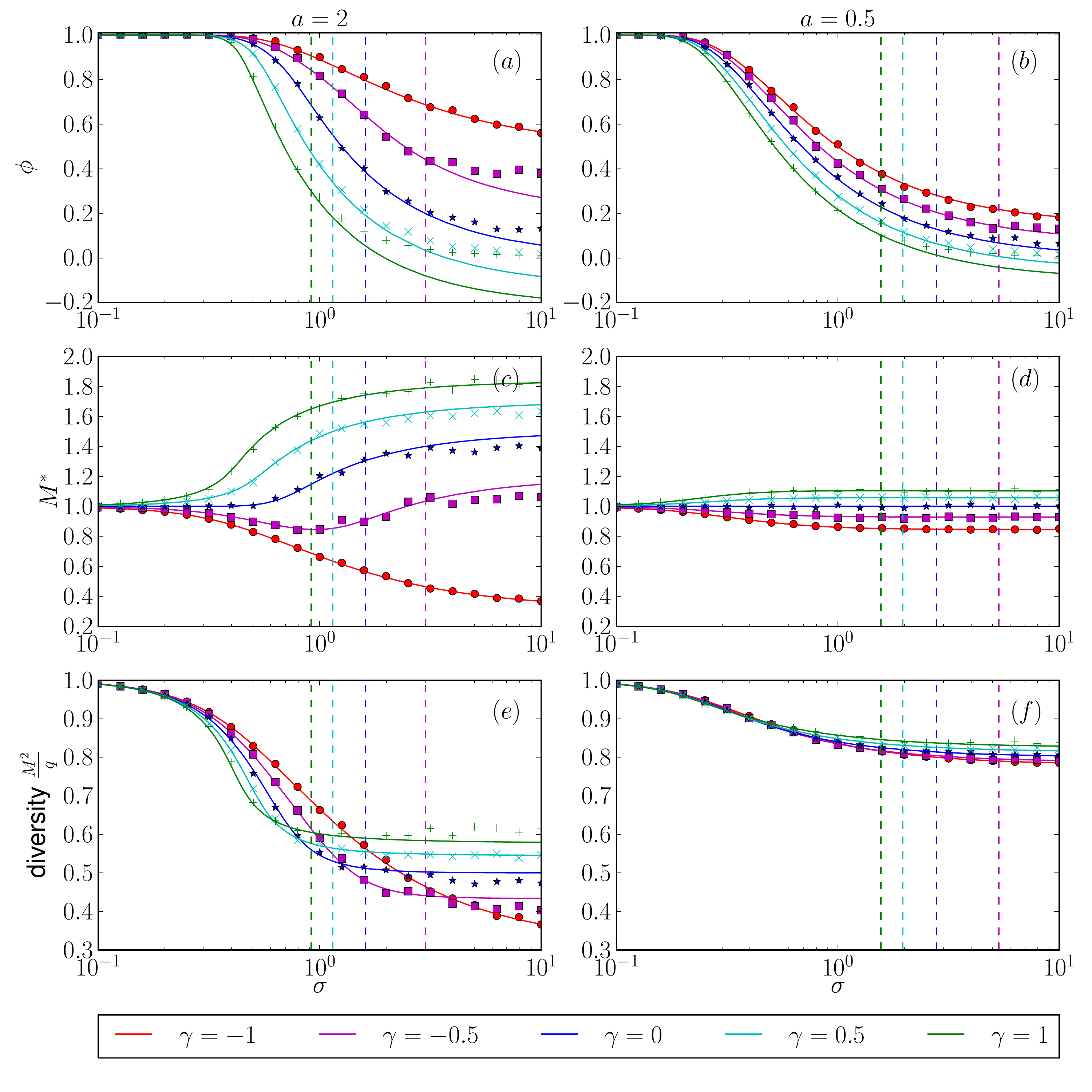}
    \caption{Comparison of theoretical predictions (lines) for the characteristic order parameters against simulations (markers). Data is for $\mu = 0$, $\gamma = -1 ~(\newmoon), -0.5~ (\blacksquare), 0~ (\bigstar), 0.5~ (\times)$, and $\gamma=1~ (+)$.
    This is for the model with piecewise linear feedback. Left-hand column [(a),(c),(e)] is for $a=2$, right-hand column [(b),(d),(f)] for $a=0.5$. The vertical dashed lines mark the onset of instability as predicted by the theory; analytical predictions can no longer be expected to match with simulations in the phase to the right of the dashed lines. The graphs show the fraction of species not saturated $\phi$, the mean abundance $M^*$, and the diversity $M^2/q$ as functions of $\sigma$.}
    \label{five2}
\end{figure}

A comparison of theory and simulation is shown in Fig.~\ref{five2}. Theoretical predictions are indicated by solid lines, results from simulations as symbols. We show the quantities $\phi$, $M^*$ and a measure of diversity related to Simpson's index. Simpson's index \cite{simpson} is the probability that two randomly chosen individuals in the community are of the same species, ${\cal S}=\sum_i \left(\frac{x_i}{\sum_jx_j}\right)^2$. For our model this index is given by ${\cal S}=q/(NM^2)$. A low value of this probability indicates high diversity of species; therefore the inverse Simpson index, ${\cal S}^{-1}=NM^2/q$ characterizes the diversity of the ecological community. The diversity scales linearly with $N$; therefore we report the relative diversity ${\cal S}^{-1}/N=M^2/q$. A more detailed explanation of how these quantities were measured from simulations and predicted from the theory can be found in Section S2 of the Supplementary Material.

The vertical dashed lines in Fig.~\ref{five2} indicate the predicted onset of instability. More precisely unique stable fixed points are predicted for small values of $\sigma$, i.e., to the left of the dashed lines. To the right of these lines the system either has multiple fixed points, or never settles down, and in either of these scenarios the analytical predictions for the stable fixed point phase can no longer be expected to apply. The figure indicates agreement between theory and simulation in the stable phase. Systematic deviations can be found in the unstable regime, although the predictions from the theory appear to remain a good approximation in some cases. Similar observations have been made in related models, see e.g. \cite{gallaasym,bunin2017}.
We note that solving Eqs.~(\ref{eq:opfp}) can lead to $z_1 > z_2$ in the unstable phase. This results in the prediction of a negative value of $\phi$ (the fraction of unsaturated species), which demonstrates further that the theory does not apply in this parameter regime.

The mean abundance, $M^*$, tends to higher values for positive values of the symmetry parameter $\gamma$, i.e., in absence of exploitative interactions and predator-prey pairs. This is shown in panels (c) and (d) in Fig.~\ref{five2}.
This can be understood from the species abundance distributions in Fig.~\ref{hist5}, and from the dependence of this distribution on $\gamma$. We find more species with larger abundances for positive $\gamma$, and more species with smaller abundances for negative $\gamma$. This is due to the dependence of $z_1$ and $z_2$ on $\gamma$ as discussed in Sec.~\ref{sec:hist}.
These effects are reduced for stronger non-linearity, as shown in Fig.~\ref{five2}(d). Generally, we find that a lower value of the saturation parameter $a$ reduces the dependence of the order parameters on $\gamma$. We also note that a much higher diversity of species is maintained for a lower saturation value [c.f. Fig.~\ref{five2}(e) and (f)].

\subsection{Onset of instability}
\begin{figure}[t!]
    \centering
    \includegraphics[width=0.85\textwidth]{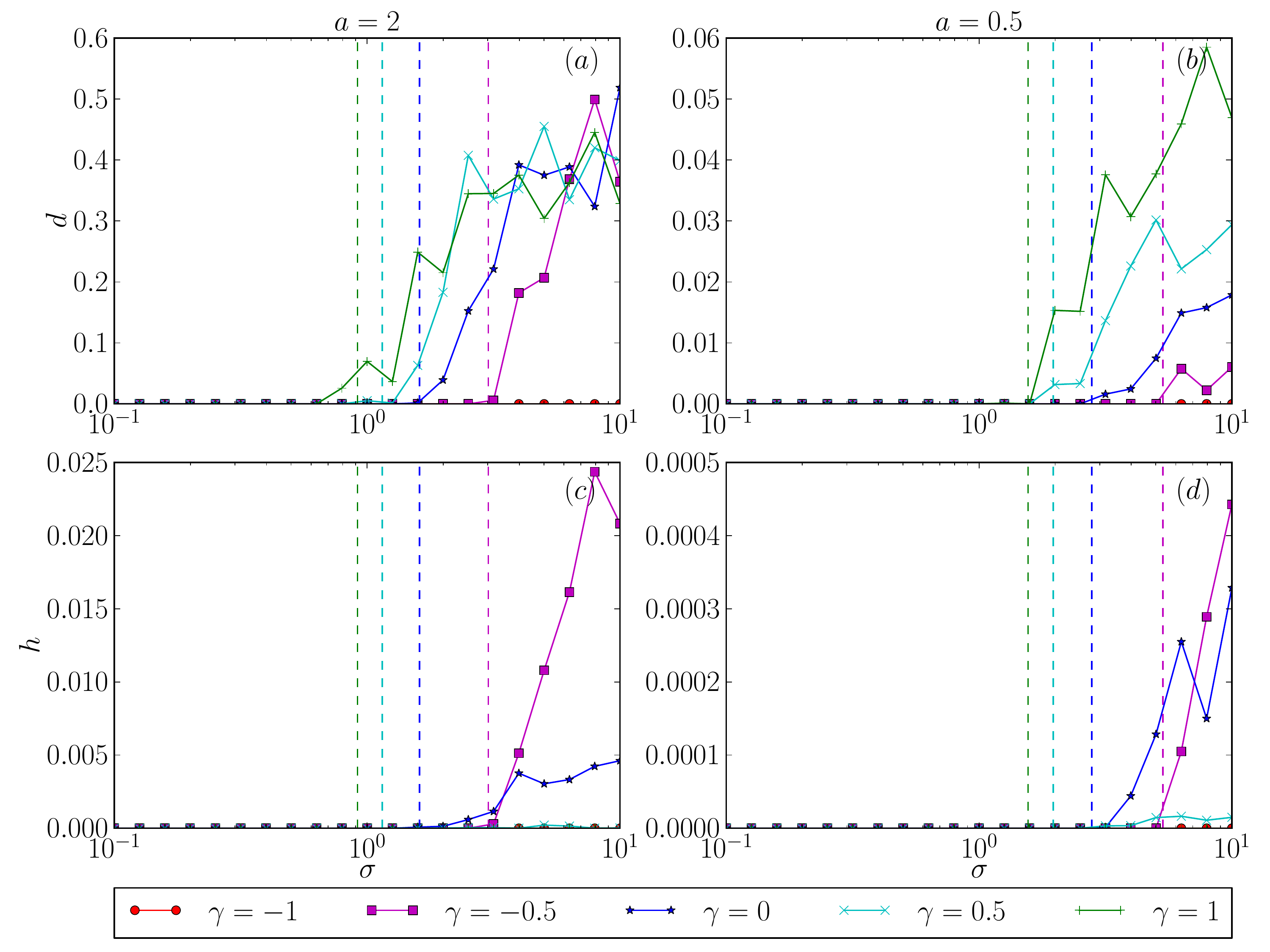}
    \caption{Onset of instability for the model with piecewise linear feedback. Left-hand column [(a),(c)] is for $a=2$, right-hand column [(b),(d)] for $a=0.5$. The vertical dashed lines mark the onset of instability as predicted by the theory. The graphs show $d$ (top) and $h$ (bottom). Data is for $\mu=0$, and $\gamma = -1 (\newmoon), -0.5 (\blacksquare), 0 (\bigstar), 0.5 (\times), 1 (+)$.}
    \label{five2dh}
\end{figure}

In Fig.~\ref{five2dh} we investigate the onset of instability in more detail. We use several indicators to detect different types of behaviour in the numerical solutions of the generalised Lotka-Volterra equations. In order to characterise the (relative) variation of species abundances over time, we calculate 
\be\label{eq:h}
h = \frac{\left<\left<x_i(t)^2\right>_{T} - \left<x_i(t)\right>_{T}^2\right>_N}{\left<\left<x_i(t)\right>_{T}^2\right>_N}.
\ee
In this expression $\avg{\cdots}_T$ indicates an average over time; this is taken in the stationary state; reported values are time averages over the last $1\%$ of trajectories (numerical integration of the generalised Lotka-Volterra equations is carried out up to final time $T_f = 200$). The notation $\avg{\dots}_N$ in Eq.~(\ref{eq:h}) denotes an average over species, $\avg{\dots}_N=N^{-1}\sum_i \cdots$.
The order parameter $h$ indicates whether or not the system settles down to a fixed point: when $h=0$ a fixed point is reached eventually, whereas positive values of $h$ indicate persistent volatile dynamics. In order to identify the phase with multiple fixed points, we have additionally run the following numerical experiments. For a fixed realisation of the interaction matrix we have generated two independent random initial conditions. We then run each of these separately, and compute the relative distance
\be
d = \frac{\left<\left< (x_i(t) - x'_i(t))^2\right>_N\right>_T}{\left<\left<x_i(t)\right>_N^2\right>_T},
\ee
where $x_i$ and $x_i'$ are the trajectories for the two sets of initial conditions. This quantity is again evaluated in the stationary state. Thus, $d\approx 0$ when the asymptotic behaviour is independent of initial conditions, and $d>0$ otherwise.

The data shown in the upper panels of Fig.~\ref{five2dh} shows that $d\approx 0$ for small heterogeneity $\sigma$ independently of the symmetry parameter $\gamma$, but that a phase with dependence on initial conditions is found when the stability threshold is crossed ($\sigma>\sigma_c$). The results in the lower panels of Fig.~\ref{five2dh} indicate that the dynamics remains volatile ($h>0$) for large values of $\sigma$ when the symmetry parameter is zero or moderately negative. The figure shows that a fixed point is almost certainly reached for $\gamma = 1$ and is likely to be reached for $\gamma = 0.5$, although these fixed points are not unique.

Comparing the scales of the left and right hand panels in Fig.~\ref{five2dh} shows that the order parameters $d$ and $h$ are much smaller for the lower value of the saturation $a$; this is due to the tighter bounding effect of the non-linear feedback.

\section{Comparison of generalised Lotka-Volterra systems with non-linear and piecewise linear feedback}\label{compare}

\subsection{Phase diagram and onset of instability}
\begin{figure}[t!]
    \centering
    \includegraphics[width=0.85\textwidth]{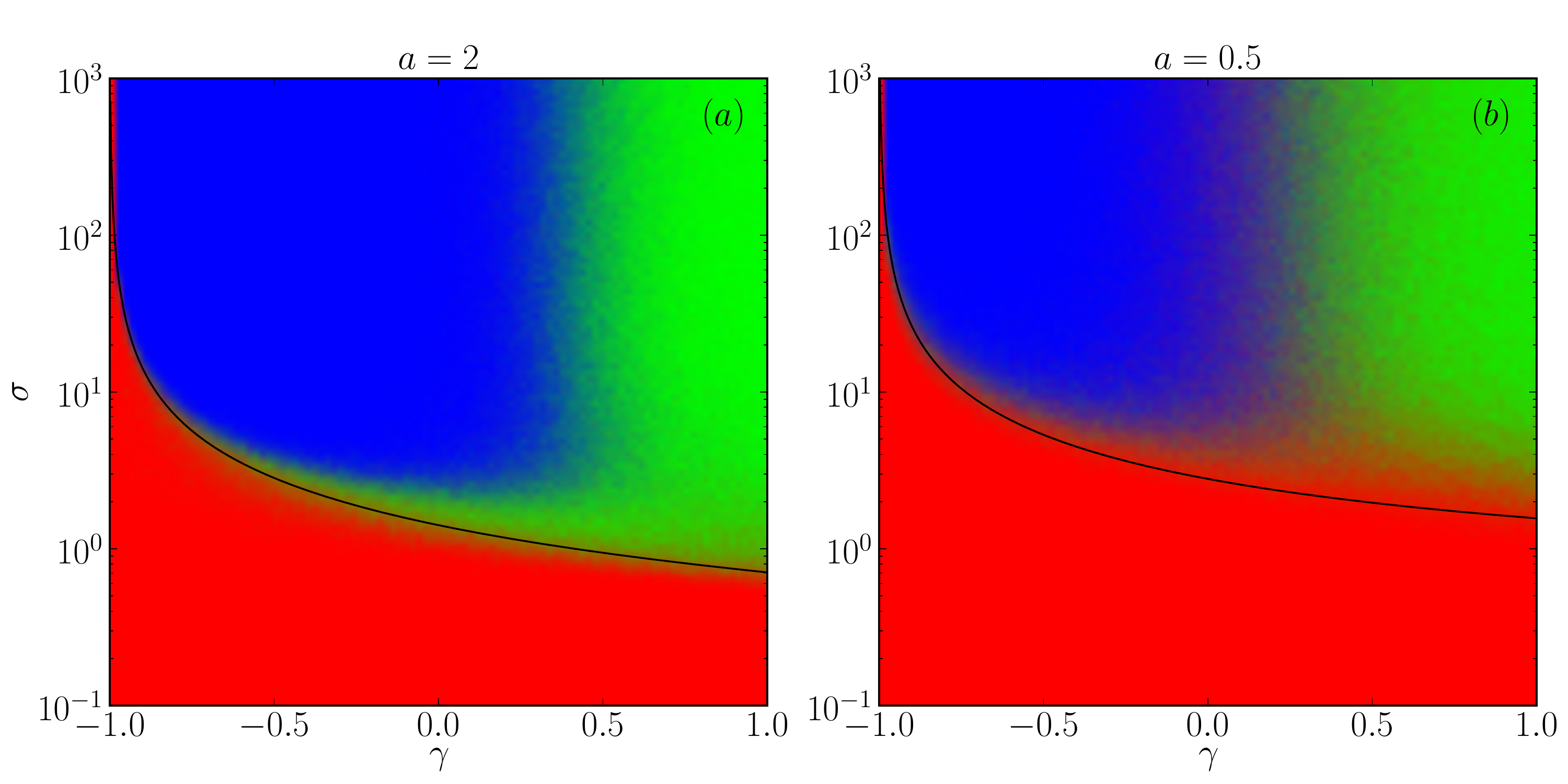}
    \caption{Phase diagram obtained from simulations of the generalised Lotka-Volterra systems with piecewise linear feedback with $N = 200$. As in Fig.~\ref{holheat} the colours (grey shading) indicate the dominant outcome in each part of parameter space [red (medium grey): unique stable fixed point; green (light grey): multiple fixed points; blue (dark grey): dynamics do not converge]. Solid black lines show the onset of instability as predicted from the generating-functional approach. Data is for $\mu = 0$, panel (a) for $a=2$, panel (b) for $a=0.5$. }
    \label{pheat}
\end{figure}

In Fig.~\ref{pheat} we show examples of the phase diagram obtained from numerical integration of the generalised Lotka-Volterra system with piecewise linear response. These are generated in the same way as in Fig.~\ref{holheat}. Red (medium grey) indicates parameter values of the phase with unique fixed points, green (light grey) indicates multiple fixed points, and blue (dark grey) indicates volatile behaviour. The black line in each panel shows the  boundary, $\sigma_c$, of the phase with a unique stable fixed point, predicted by the theory. As seen in the figure the theory is in agreement with results from numerical integration of the generalised Lotka-Volterra system. We attribute remaining minor discrepancies to finite integration time, finite time steps, and finite species number.

Comparing Figs.~\ref{holheat} and \ref{pheat}, we find that the behaviour of the systems with non-linear and piecewise linear feedback are very similar. Unique fixed points are reached for values of $\sigma$ below a critical point for all values of the symmetry parameter $\gamma$, with much higher critical values $\sigma_c$ for lower $\gamma$. This indicates higher stability for asymmetric couplings than in the symmetric case. Above this critical value of the heterogeneity parameter we find multiple fixed points for positively correlated interactions, and persistent volatile behavior, such as limit cycles, chaos, or potentially heteroclinic cycles for negatively correlated couplings. In both Fig.~\ref{holheat} and Fig.~\ref{pheat}, one notices the stabilising effect of a lower value of the saturation parameter $a$ in the non-linear feedback, i.e., for smaller $a$ one finds a larger red (medium grey) area indicating stable unique fixed points, and higher critical values for $\sigma$.

We have observed that the area of unique fixed points in the phase diagrams becomes bigger for smaller values of the number of species $N$. Conversely, we would expect the results from simulations to converge to the analytical prediction in Fig.~\ref{pheat}(b) for higher values of $N$. Interestingly, this effect is more pronounced for $a=0.5$ than for $a=2$. I.e. for $a=0.5$ the system retains more structural stability as $N$ is increased than for $a=2$. This underlines the stabilising effect of the non-linear feedback.

 \begin{figure}[t]
    \centering
    \includegraphics[width=0.85\textwidth]{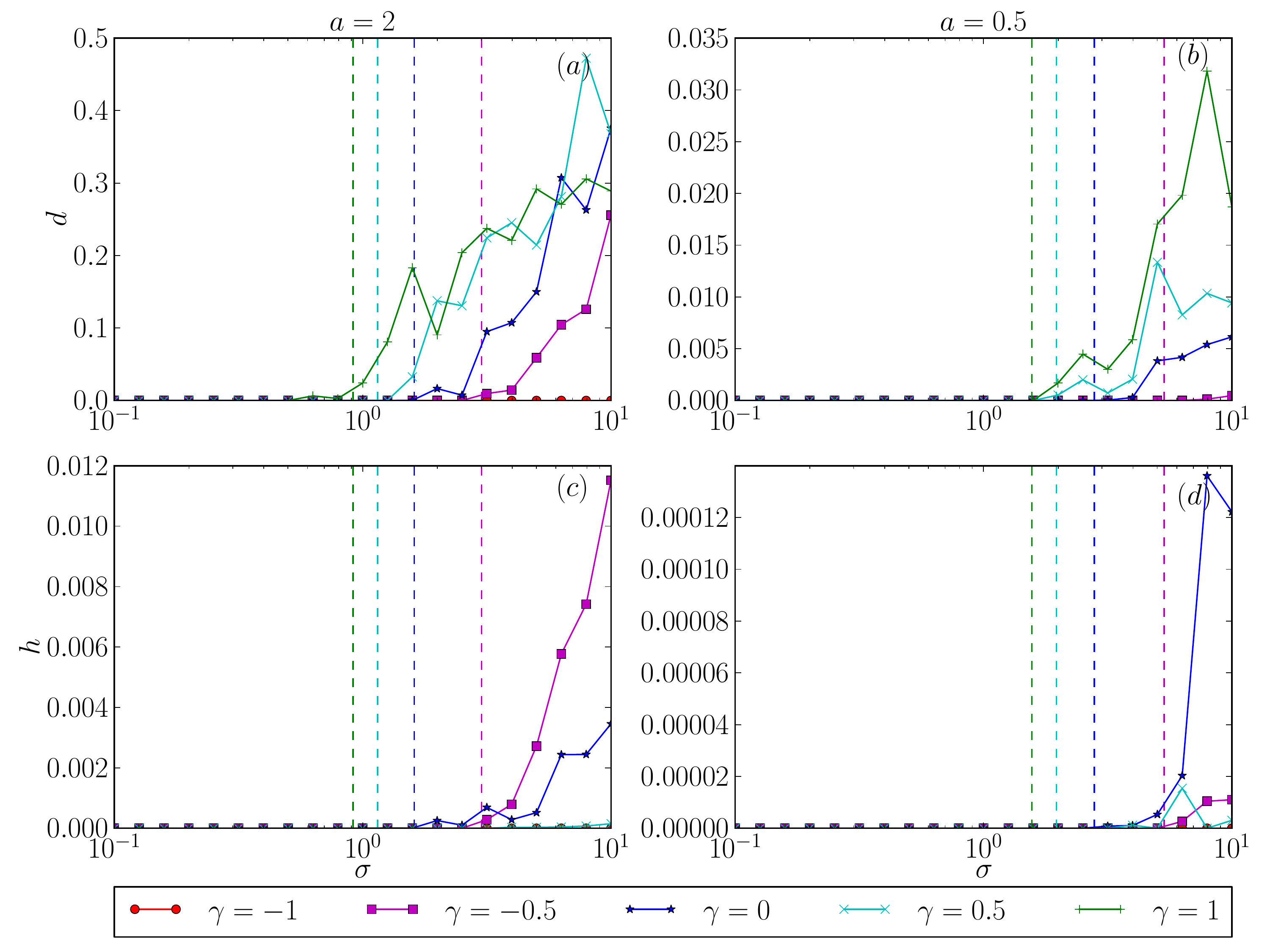}
    \caption{Onset of instability for non-linear feedback. Left-hand column [(a),(c)] is for $a = 2$, right-hand column [(b),(d)] for $a = 0.5$. The vertical dashed lines mark the onset of instability predicted for the piecewise feedback function from theory. The graphs show $d$ (top) and $h$ (bottom). Data is for $\mu=0$, $\gamma = -1 (\newmoon), -0.5 (\blacksquare), 0 (\bigstar), 0.5 (\times), 1 (+)$.}
    \label{five3dh}
\end{figure}

In order to make the comparison between the two models more precise we report results for the order parameters $h$ and $d$ from numerical simulations of the model with non-linear feedback in Fig.~\ref{five3dh}, along the analytical prediction for the onset of instability in the model with piecewise linear response. The data shows that the system with non-linear feedback has very similar behaviour as that with the piecewise linear feedback. We find volatile dynamics for anti-correlated interactions past the critical interaction heterogeneity, and multiple fixed points for correlated interactions. The point at which $h$ and $d$ become non-zero is very close to the onset of instability predicted by the theory for the model with piecewise linear feedback.

\subsection{Order parameters in the stable phase}

\begin{figure}[t!]
    \centering
    \includegraphics[width=0.85\textwidth]{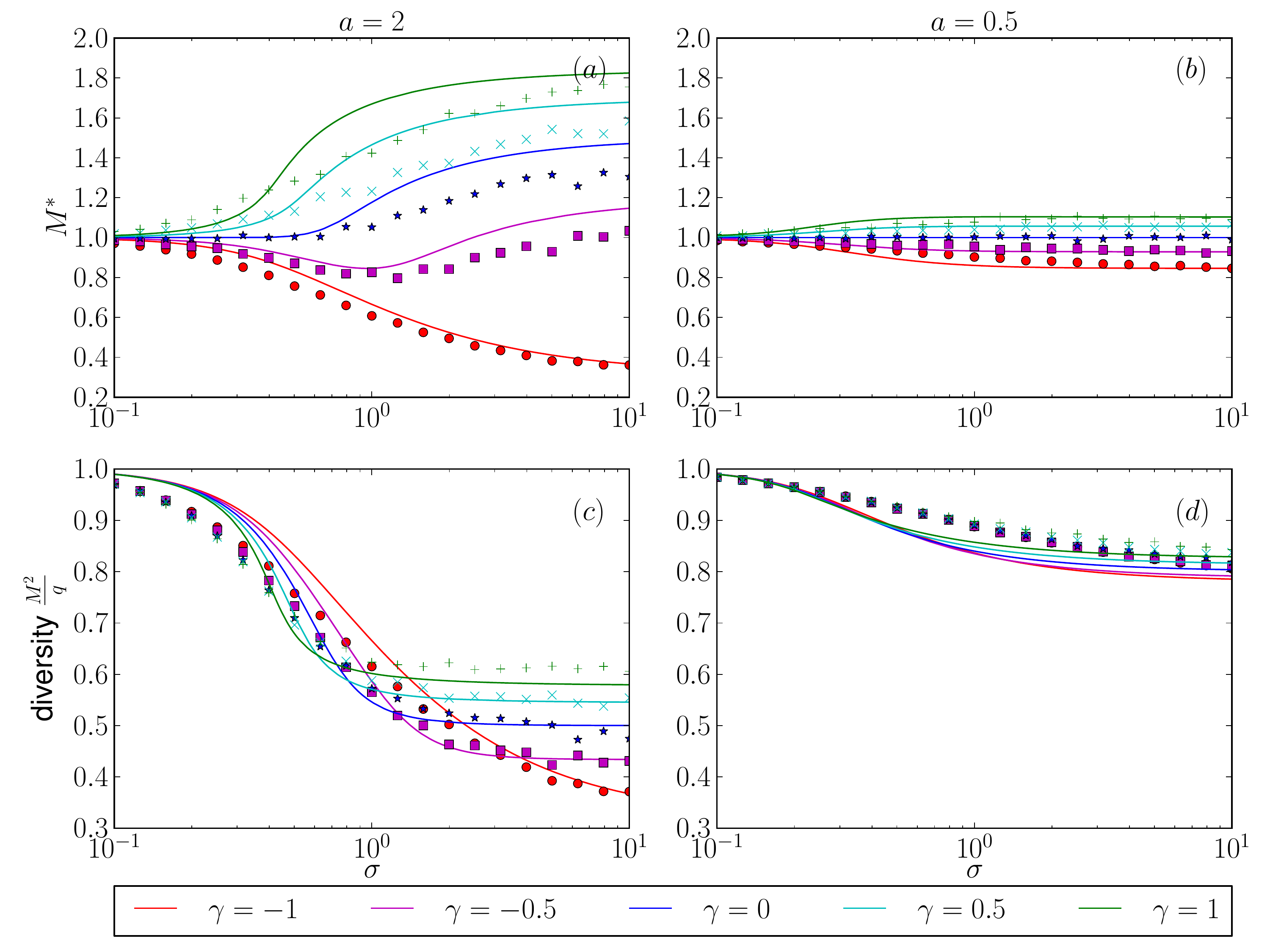}
    \caption{Comparison of theoretical predictions for $M^*$ and diversity for the piecewise feedback (lines) against simulations for non-linear feedback (markers). Left-hand column [(a),(c)] is for $a=2$, right-hand column [(b),(d)] for $a=0.5$. Data is for $\mu=0$, $\gamma = -1 (\newmoon), -0.5 (\blacksquare), 0 (\bigstar), 0.5 (\times), 1 (+)$.}
    \label{five3}
\end{figure}

In Fig.~\ref{five3} we compare results from numerical integration of the generalised Lotka-Volterra equations with non-linear feedback (markers), with the analytical solutions for piecewise linear feedback (lines). As seen in the figure the general behaviour of the mean abundance $M^*$ and diversity as functions of $\sigma$ and $\gamma$ are similar in both models.

The main difference between the piecewise linear function and the non-linear function is how they approach their upper and lower limits $\pm a$. The piecewise linear function $g_P(u)$ approaches its limits linearly, and attains them at $u=\pm a$ [$g_P(\pm a) = \pm a$]; the non-linear Hill function $g_H$ approaches the limits much more slowly, and only attains them asymptotically. As a consequence, we have $|g_H(u)|<|g_P(u)|$ for $|u|>a/2$.
The differences in the two functions account for the differences we see between the results in Fig.~\ref{five3}. We find the order parameters $M^*$ and diversity to display a much smoother dependence on heterogeneity for the non-linear function (markers) than for the piecewise function.

For the larger values of $\sigma$ shown in Fig.~\ref{five3}(a), we find that $M^*$ is lower for the non-linear function for all values of $\gamma$.
For $a=2$ species die before they can saturate to the lower boundary. In this case the difference in saturation is therefore only present at the upper limit, and this results in lower values of $M^*$ for the non-linear function.
This is because species are closer to upper saturation point for higher values of $\sigma$, where the non-linear function is lower in magnitude than the piecewise linear function. The lower value of the non-linear function causes these species to have lower abundances than in the piecewise case.

For $a=0.5$, species are able to reach both saturation points before they can die, therefore the difference in the two feedback functions affects the species both at the upper and lower boundaries.
As a consequence, we do not see the same consistent effect of lower $M^*$ as we did in the case of $a=2$, see Fig.~\ref{five3}(b). 

In the limit of infinite $\sigma$, both the non-linear and piecewise linear function are effectively equivalent. In this limit all species abundances will be saturated at either boundary, $x_i = 1+a$ or $x_i = (1-a)\Theta(1-a)$. The fraction of species saturated at each boundary is the same for either function, for a given $\gamma$. This results in the same limiting values for the order parameters $M^*$ and diversity in both models.

\section{Dependence of stable region on model parameters}\label{sec:dependence}

In the previous section we compared results from analysis and simulations of the piecewise linear feedback to demonstrate that our theory correctly predicts the nature of the system in the regime of unique fixed points, see Fig.~\ref{five2}. The theory also correctly predicts the critical value of $\sigma$ where this regime ends, as demonstrated in Fig.~\ref{five2dh}.
We then compared predictions for the model with piecewise linear feedback with simulation results for the model with non-linear feedback. We found a similar general dependence of the system's order parameters on $\sigma$, $\gamma$, and $a$, see Fig.~\ref{five3}. The onset of instability, $\sigma_c$ is also very similar on both models, as shown in Fig.~\ref{five3dh}. We conclude that the predictions of our theory for the piecewise linear feedback function are a good approximation to the behaviour of the model with non-linear feedback. It is therefore appropriate to use the theory we have developed to investigate further how the stability of the ecosystem with saturating non-linear feedback depends on the key model parameters.

\subsection{Dependence of stability on the saturation parameter $a$}

\begin{figure}[t!]
    \centering
    \includegraphics[width=0.65\textwidth]{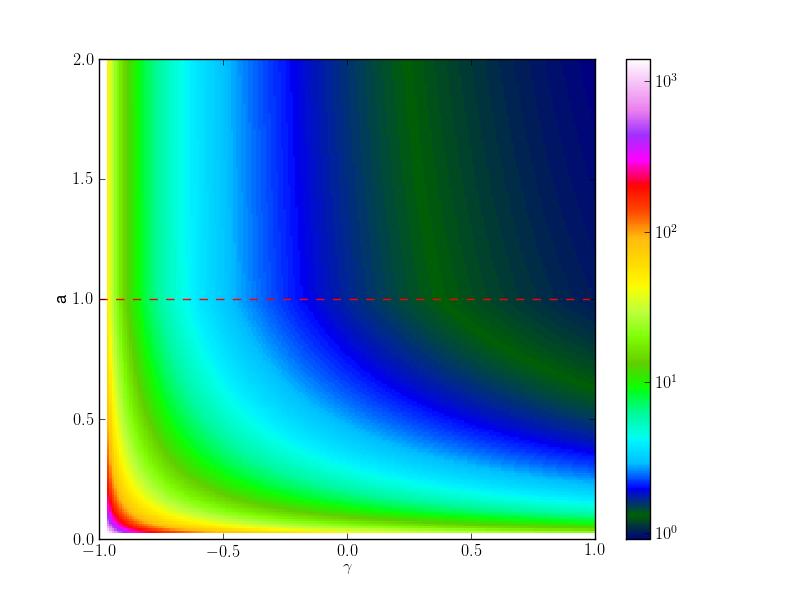}
    \caption{Critical value of the heterogeneity parameter, $\sigma_c$, plotted as a colour map in the $a - \gamma$ plane at fixed $\mu = 0$. Higher values of $\sigma_c$ indicate higher stability. The dashed line indicates $a=1$ (saturation parameter equal to carrying capacity), see text for further discussion.}
    \label{acrit}
\end{figure}

We have so far shown results only for $a = 2$ and $a=0.5$. These fall on either side of the  carrying capacity which was set to one. In Fig.~\ref{acrit} we provide a more general picture, and show how the critical value of the heterogeneity, $\sigma_c$, depends on the saturation parameter $a$, and on the symmetry parameter $\gamma$. In this figure we fix the mean value $\mu = 0$ of the interaction matrix elements.

As one would expect, the range of the stable region increases as the non-linear feedback becomes more restricted (i.e., as $a$ is lowered). This effect is particular relevant when the saturation parameter is lower than the carrying capacity (i.e, for $a<1$). We note that in this regime the critical strength of the heterogeneity is a decreasing function of both $a$ and $\gamma$. For $\gamma < 0$ we also note that the stability has a similar dependence on the two parameters $a$ and $\gamma$, which is demonstrated by the symmetry in the bottom left quadrant of Fig.~\ref{acrit}.
If the saturation parameter $a$ exceeds the carrying capacity ($a>1$), its influence on the size of the stable region is small.

\subsection{Dependence of stability on cooperation parameter $\mu$}

\begin{figure}[t!]
    \centering
    \includegraphics[width=\textwidth]{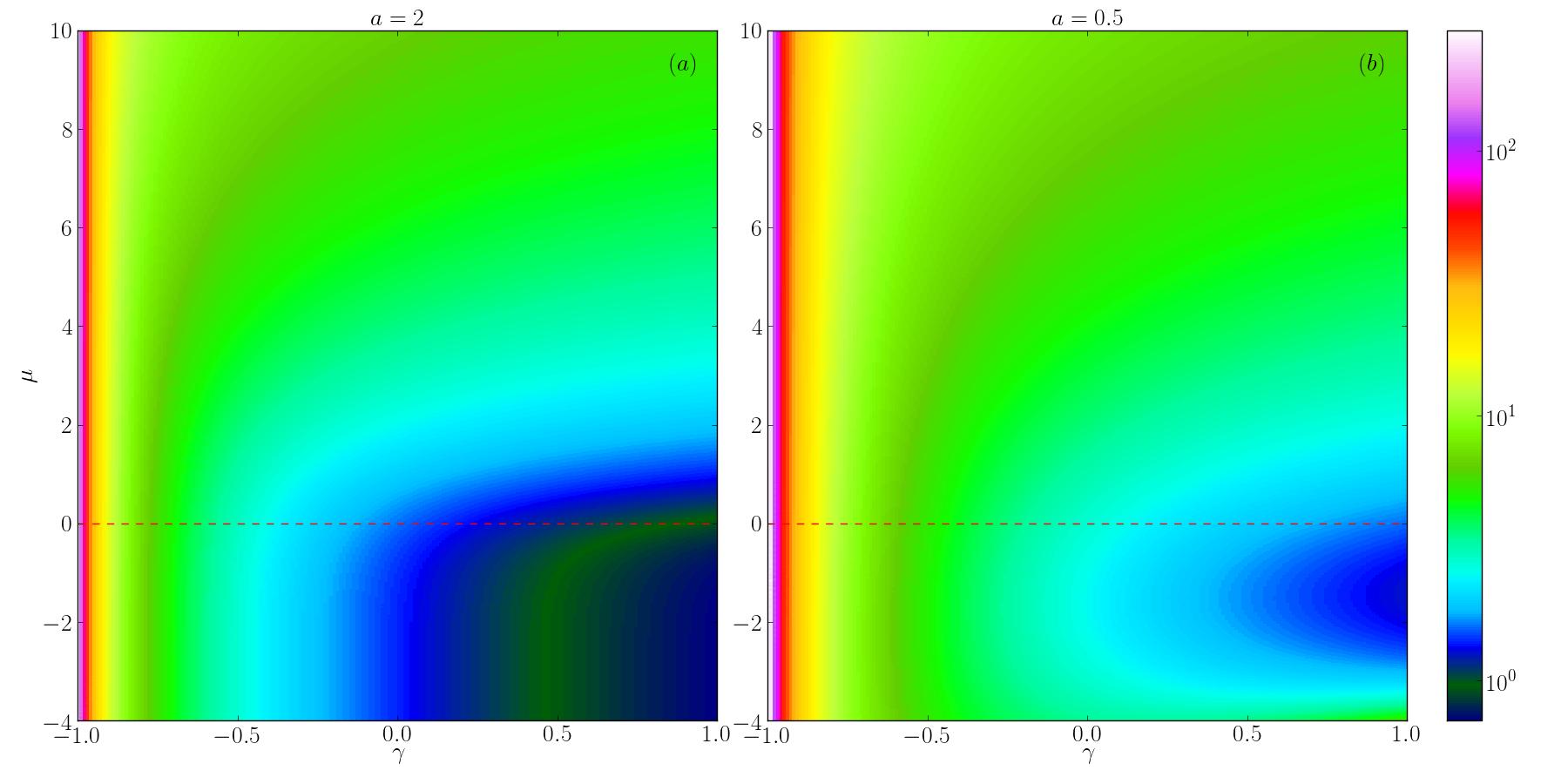}
    \caption{Critical $\sigma_c$ plotted as a colour map in the $\mu - \gamma$ plane at fixed $a = 2$ (left) and $a=0.5$ (right). Higher values of $\sigma_c$ indicate higher stability. The data along the dashed line indicates values of $\sigma_c$ for $\mu=0$, these are as previously given in the black lines of Fig.~\ref{pheat}. }
    \label{mcrit}
\end{figure}

\begin{figure}[t!]
    \centering
    \includegraphics[width=\textwidth]{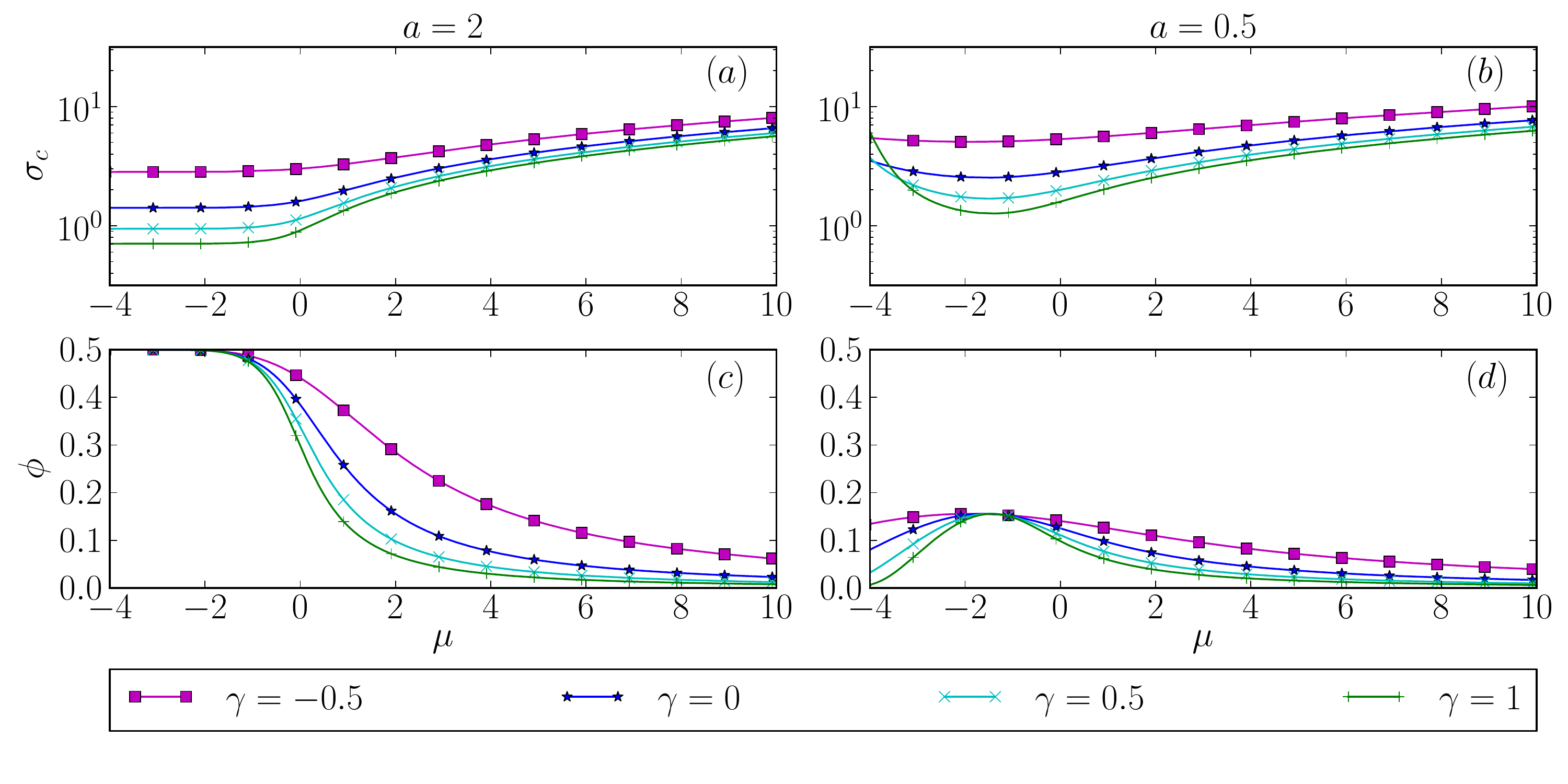}
    \caption{Location of the onset of instability $\sigma_c$, and fraction of unsaturated species, $\phi$, for varying $\mu$ at $a = 2$ (left) an $a = 0.5$ (right). Results in the figure are from the theory, the different values for $\gamma$ are indicated by different symbols [$\gamma = -0.5 (\blacksquare), 0 (\bigstar), 0.5 (\times), 1 (+)$].}
    \label{mhist}
\end{figure}

We have not yet considered how stability varies with the co-operation parameter $\mu$, i.e., the mean interacting strength between species. In the context of gut bacteria it has been argued that increasing the co-operation between species of an ecological system can reduce the system's stability \cite{kat}, with higher stability found for more competitive systems.

Previous theoretical studies \cite{bunin2016,bunin2017,gallaepl} of  generalised random Lotka-Volterra systems with linear feedback have found $\sigma_c$ to be independent of $\mu$ so long as $\mu \leq 0$. We note that the interaction term between the species carries the opposite sign in \cite{bunin2016, bunin2017} relative to our notation, implying opposite sign conventions in particular for the parameter $\mu$. In \cite{bunin2016, bunin2017} a second critical value of the heterogeneity is found; if the strength of the heterogeneity exceeds this value, the system displays unbounded growth. This value is found to depend on $\mu$, and to be equal to zero for $\mu \geq 1$; that is to say, when $\mu>1$ the random generalised Lotka-Volterra system with linear feedback always exhibits unbounded growth regardless of the amount of heterogeneity.  

The saturated non-linear feedback in our model causes the abundances to be constrained to the interval $1-a\Theta(1-a) \leq x_i(t) \leq 1+a$, and as a consequence the system cannot display unbounded growth. The critical value for $\sigma$ however, is now dependent on the value of $\mu$. In Fig.~\ref{mcrit} we show how $\sigma_c$ varies with $\mu$ and $\gamma$ for $a=2$ and $a=0.5$; these results are from numerical evaluation of the self-consistency equations obtained from the generating functional analysis.

For $a = 0.5$, we find a minimum value for $\sigma_c$ as a function of $\mu$, see the lower right corner of Fig.~\ref{mcrit}(b). This minimum value corresponds to a maximum proportion of unsaturated species $\phi$, this is shown in Sec.~S1~J and demonstrated in Fig.~\ref{mhist}. For values of $\mu$ away from this extremal point, $\phi$ decreases as more species become saturated at either $1+a$ (increasing $\mu$) or $1-a$ (decreasing $\mu$), this in turn gives the system a higher stability.
For $a=2$ we also find increasing stability for higher values of $\mu$. However, we do not find a minimum point for stability (as a function of $\mu$). Instead $\sigma_c$ monotonically decreases with decreasing $\mu$, tending to a constant. This is because the fraction of species at the lower saturation point ($x^* = 0$ for $a=2$) does not increase past $0.5$ as $\mu$ is decreased. 

The transition line shown in the $\mu - \sigma$ plane shown in Fig.~\ref{mhist} (a) and (b) can be compared to the phase diagrams for linear feedback in \cite{bunin2016, bunin2017, felixroy}. For linear feedback the transition line is not dependent on $\mu$ as it is in the case of non-linear feedback.
These results are different to those found in \cite{kat}, where a model with linear feedback was used, co-operation then does not promote stability. For saturating non-linear feedback we find, instead, that stability is increased for higher values of the co-operation parameter $\mu$.

\section{Conclusions}\label{sec:concl}
In summary we have analysed generalised Lotka-Volterra communities with random interactions and non-linear feedback. Specifically, we have studied  systems in which the total feedback on the growth rate of any one species saturates via a non-linear function. Simulations of such systems reveal three different types of behaviour. When the variation in interaction coefficients is small, convergence to stable fixed points is found for a wide range of values of the remaining model parameters. These fixed points are found to be unique, in the sense that the asymptotic composition of resulting community does not depend on the starting point of the dynamics for any realisation of the random interaction matrix. This globally stable behaviour is found provided the heterogeneity of interactions $\sigma$ does not exceed a critical value. This critical value in turn depends on the combination of the parameter $a$ indicating saturation value, the co-operation parameter $\mu$, and the symmetry parameter $\gamma$. Above the critical interaction heterogeneity two types of behaviours are observed: when the heterogeneity in interaction coefficients is such that it promotes symmetric interactions we observe that the generalised Lotka-Volterra dynamics can have multiple stable fixed points, and which one is reached asymptotically depends on the choice of initial conditions. While we often find fixed points in simulations in this phase, the system could -- in principle -- also exhibit heteroclinic cycles for larger numbers of species $N$. For negatively correlated heterogeneity instead we typically observe persistently volatile dynamics.
 
The critical value for $\sigma$ is much higher for a lower value for $\gamma$, with $\sigma_c \to \infty$ for $\gamma = -1$, when all interactions are exploitative. Low values of the saturation parameter (i.e., strongly non-linear feedback) increases the critical $\sigma$, i.e., the range of global stability is larger. This effect is seen in particular when the saturation value $a$ is smaller than the carrying capacity. Previous studies have found that stability decreases with an increasing co-operation parameter $\mu$ \cite{kat,bunin2016,bunin2017} in models with linear feedback. 
In this linear case the interaction with other species is not bounded, and a high degree of cooperation can lead to unlimited growth. With non-linear feedback the growth of abundances is bounded, and co-operation no longer leads to unlimited growth. With non-linear feedback we have found that increasing the co-operation parameter $\mu$ can increase stability, even when the feedback is linear over a wide range (i.e., when $a$ is large); the key factor for stability is saturation at very large or very small arguments.

We conclude by briefly speculating about potential biological implications of our findings. The human gut has evolved with the microbiome, and has adaptated to promote to stability as this is important for good health. Our findings suggest, that one effective way to increase the system's global stability is to decrease $\gamma$, i.e., to exhibit more exploitative predator-prey like interactions. It may be difficult for the human gut to have an influence over the specific interactions types present between the species, as these will be a function of the microbes themselves. The human gut has however, been found to promote ecosystem stability by host feeding, immune suppression and spatial structure \cite{kat} \cite{leash}. Spatial structure has the effect of reducing the interaction strength between populations of different species, which has the effect of reducing $\sigma$, which our model has shown to increase global stability. Host feeding and immune suppression may work by enforcing a bound on species population size from above and below, which may result in a similar effect produced by non-linear feedback explored in this paper. We have shown that enforcing a tight population bound (low $a$) promotes ecosystem global stability, but also results in a higher system diversity (Fig.~\ref{five2}) which is also beneficial for health. Whilst employing non-linear feedback, the cooperation within the system (parametrised by $\mu$) can be increased without the adverse effect of destabilising the system \cite{kat}. This allows a more cooperative and efficient microbiome without compromising stability. Non-linear feedback has been observed in other ecosystems \cite{leslie,holling1,holling2} and one would expect evolution to utilize this beneficial effect in the gut.

\acknowledgements{We thank G. Biroli and F. Roy for discussions. LS acknowledges a PhD studentship by the Engineering and Physical Sciences Research Council (EPSRC UK), grant number EP/N509565/1. Partial financial support has been received from the Agencia Estatal de Investigacion (AEI, Spain) and Fondo Europeo de Desarrollo Regional (FEDER, EU) under Project PACSS (RTI2018-093732-B-C22), and the Maria de Maeztu Program for Units of Excellence in R\&D (MDM-2017-0711).}

\clearpage
 \onecolumngrid
 \renewcommand{\theequation}{S\arabic{equation}}
 \setcounter{equation}{0}
  \renewcommand{\thesection}{S\arabic{section}}
 \setcounter{section}{0}
 
   \renewcommand{\thefigure}{S\arabic{figure}}
 \setcounter{figure}{0}

   \renewcommand{\thepage}{S\arabic{page}}
 \setcounter{page}{1}

\begin{center}
{\bf Ecological communities from random generalised Lotka-Volterra dynamics with non-linear feedback} \\
~\\

{\bf Laura Sidhom and Tobias Galla}\\
\medskip
~\\
{\bf \Large---~Supplementary Material~---}

 \end{center}
 
 \label{supp}

\section{Details of the generating-functional analysis}
The generating-functional method \cite{dominicis} is a well-established technique in the theory of disordered systems. It is used to study spin glasses, neural networks and other systems with quenched disorder. The literature on the method is extensive, sources can for example be found in \cite{mpv, coolendyn}. Recent reviews include \cite{sollich}. The method was first applied to random replicator systems in \cite{opper}. A detailed account of the steps required to carry out the calculation for Lotka-Volterra models with linear feedback can be found in \cite{gallaepl2018}. We note that the resulting dynamical mean-field theory (an effective process for a representative species) can also be derived using alternative methods, see e.g. \cite{bunin2017} or \cite{felixroy}. In this Section we describe the calculation in the context of the generalised Lotka-Volterra model with non-linear feedback. For a more general account of the method and further background we refer the reader to the sources cited earlier in this paragraph.

\subsection{Expression for the generating functional}
The generalised Lotka-Volterra equations can be written as
\begin{align}
\frac{\dot{x}_i(t)}{x_i(t)} = 1-x_i(t) + g\left(\sum_j\alpha_{ij}x_j(t)+h_i(t)\right) ,
\end{align}
where we have introduced $h_i(t)$ to generate response functions and will be later set to zero.

We use this expression to set up the generating functional for the process,
\BE 
Z[\pmb{\psi}]  &=& \int D[\pmb{x}]
\exp \left( i \sum_i \int \psi_i(t)x_i(t) dt \right) \left[ \prod_i  p(x_i(0))\right] \nonumber \\
&&\times \prod_{i,t} \delta \left\{ \frac{\dot{x}_i(t)}{x_i(t)} - \left[1-x_i(t) + g\left(\sum_j\alpha_{ij}x_j(t)+h_i(t)\right)  \right]\right\}.
\EE
The integral $\int D[\pmb{x}]=\prod_i\prod_t \int dx_i(t)$ is over all possible paths and random initial conditions; we assume that the latter factorise over the different species, $i$. 

We can write the delta function as a Fourier transform to give
\begin{align}
Z[\pmb{\psi}] &= \int  D\hat{\pmb{x}}D\pmb{x} \left[ \prod_i p(x_i(0))\right]  \exp \left( i \sum_i \int \psi_i(t)x_i(t) dt \right)
\nonumber \\
&\times \exp \left\{ i \int \sum_i \hat{x}_i(t)  \left[ \frac{\dot{x}_i(t)}{x_i(t)} -1 +x_i(t) - g\left(\sum_j\alpha_{ij}x_j(t)+h_i(t)\right)  \right] dt \right\},
\end{align}
where factors of $2\pi$ have been absorbed in the measure via the definition
\begin{align} D\hat{\pmb{x}}D\pmb{x} \equiv \prod_i\prod_t  \frac{d\hat{x}_i(t)dx_i(t)}{2\pi}\end{align} 

\subsection{Auxiliary variable for disordered argument inside the functional feedback}
In order to be able to carry out the disorder average at a later point, we isolate the terms containing the random coupling matrix by introducing the auxiliary variables
\begin{align}
f_i(t) = \sum_j \alpha_{ij}x_j(t) + h_i(t).
\end{align}
In practical terms we do this by writing unity in the following form,
\be
\int \delta\big[f_i(t) - \sum_j \alpha_{ij}x_j(t) - h_i(t)\big] ~ df_i(t)=1,
\ee
for all $i$ and $t$. We then have
\be\label{expdelta}
 \int  D\hat{\mathbf{f}}D\mathbf{f} \exp\left[ i\sum_i \int \hat{f}_i(t)\left( f_i(t) - \sum_j \alpha_{ij}x_j(t) - h_i(t)\right) dt \right]=1.
\ee
Inserting this into the generating functional, we find
\begin{align}
Z[\boldpsi] &= \int  p(\mathbf{x}(0)) \exp \left( i \sum_i \int \psi_i(t)x_i(t) dt \right)
\exp \left[ i \int \sum_i \hat{x}_i(t)  \left( \frac{\dot{x}_i(t)}{x_i(t)} -1 +x_i(t) - g\left(f_i(t)\right) \right) dt \right] \nonumber \\
&\times \exp\left[i \int \sum_i \hat{f}_i(t)\left( f_i(t) - \sum_j \alpha_{ij}x_j(t) - h_i(t) \right)dt\right]D\mathbf{f}D\hat{\mathbf{f}} D\mathbf{x}D\hat{\mathbf{x}}.
\end{align}
For later purposes we introduce the average
\BE
&&\avg{F(\mathbf{f},\hat{\mathbf{f}},\mathbf{x},\hat{\mathbf{x}})} \nonumber \\
&=&\int p(\mathbf{x}(0)) \exp \left[ i \int \sum_i \hat{x}_i(t)  \left( \frac{\dot{x}_i(t)}{x_i(t)} -1 +x_i(t) - g\left(f_i(t)\right) \right) dt \right] \nonumber \\
&&\times \exp\left[i \int \sum_i \hat{f}_i(t)\left( f_i(t) - \sum_j \alpha_{ij}x_j(t) - h_i(t) \right)dt\right] F(\mathbf{f},\hat{\mathbf{f}},\mathbf{x},\hat{\mathbf{x}})~D\mathbf{f}D\hat{\mathbf{f}} D\mathbf{x}D\hat{\mathbf{x}},
\EE
for functions (or functionals) $F(\mathbf{f},\hat{\mathbf{f}},\mathbf{x},\hat{\mathbf{x}})$.

\subsection{Calculation of moments from the generating functional}
Moments of the dynamical variables are obtained as derivatives of the generating functional. For example, we have 
\begin{align}\label{diff1}
\left. \pdv{Z[\boldpsi]}{\psi_i(t)} \right|_{\boldpsi = \mathbf{0}} = i\left<x_i(t)\right>,
\end{align}
as well as
\begin{align} \label{diff2}
\left.\pdv{Z[\boldpsi]}{\psi_i(t)}{\psi_i(t')} \right|_{\boldpsi = \mathbf{0}} = -\left<x_i(t)x_i(t')\right>.
\end{align}
Response functions can be obtained as follows,
\begin{align}\label{diff3}
\left.\pdv{Z[\boldpsi]}{\psi_i(t)}{h_i(t')}\right|_{\boldpsi = \mathbf{0}} = i\pdv{\left<x_i(t)\right>}{h_i(t')} = \left<x_i(t)\hat{f}_i(t')\right>.
\end{align}

It is also useful to note that $Z[\boldpsi = \mathbf{0}] = 1$ for all choices of the perturbation fields $\{h_i(t)\}$. This indicates that
\begin{align}\label{diff4}
\pdv{Z[\boldpsi = \mathbf{0}]}{h_i(t)} = -i\left<\hat{f}_i(t)\right> = 0,
\end{align}
and
\begin{align}\label{diff5}
\pdv{Z[\boldpsi = \mathbf{0}]}{h_i(t)}{h_i(t')} = -\left<\hat{f}_i(t)\hat{f}_i(t')\right> = 0.
\end{align}

\subsection{Disorder average}
We now carry out the average over the disorder. The Gaussian random matrix elements $\alpha_{ij}$ have moments
\begin{align}
\overline{\alpha_{ij}} = \frac{\mu}{N}, \quad \quad \overline{\alpha_{ij}^2}-\frac{\mu^2}{N^2} = \frac{\sigma^2}{N}, \quad \quad \overline{\alpha_{ij}\alpha_{ji}} - \frac{\mu^2}{N^2} = \frac{\gamma\sigma^2}{N}.
\end{align}
Focusing on the term in the generating functional containing the disorder, and noting that $\alpha_{ii} = 0 \quad \forall i$, we find
\BE
&&\overline{\exp\left(-i\int \sum_{i \neq j} \hat{f}_i(t) \alpha_{ij}x_j(t) dt\right)} \nonumber \\
&=& \exp\left(-\mu N\int F(t) M(t) dt - \frac{\sigma^2 N}{2}\int\{ L(t,t')C(t,t') + \gamma K(t,t')K(t',t)\} dtdt' \right),
\EE
where we have introduced the macroscopic variables
\BE
C(t,t') &=& \frac{1}{N} \sum_i x_i(t)x_i(t'), \nonumber \\
 L(t,t') &=& \frac{1}{N} \sum_i \hat{f}_i(t)\hat{f}_i(t'),\nonumber \\
 K(t,t') &=& \frac{1}{N} \sum_i x_i(t)\hat{f}_i(t'),\nonumber \\
 M(t)& =& \frac{1}{N} \sum_i x_i(t),\nonumber \\
 F(t)& =&  \frac{i}{N} \sum_i \hat{f}_i(t).
\EE
It is also useful to introduce
\begin{align}
J(t) = \frac{i}{N} \sum_i \hat{f}_i(t)f_i(t).
\end{align}
Inserting suitable delta-functions in their exponential representation similar to Eq.~(\ref{expdelta}) into the generating functional, we find

\begin{align}\label{eq:zbar}
\overline{Z[\mathbf{\psi}]} = \int \exp\left[N\left(\Omega + \Phi + \Psi \right) \right] D\mathbf{C}D\mathbf{\hat C}D\mathbf{L}D\mathbf{\hat L}
D\mathbf{K}D\mathbf{\hat K}D\mathbf{M}D\mathbf{\hat M}D\mathbf{F}D\mathbf{\hat F}D\mathbf{J}D\mathbf{\hat J},
\end{align}
with
\begin{align}\label{eq:omega}
\Omega &= \frac{1}{N} \sum_i \ln \left\{ \int p(x_i(0)) \exp \left( i \int \psi_i(t)x_i(t) dt \right)\exp \left( -i \int \hat{f}_i(t) h_i(t) dt \right)\right. \nonumber \\
&\times \exp \left[ i \int \hat{x}_i(t)  \left( \frac{\dot{x}_i(t)}{x_i(t)} -1 +x_i(t) - g\left(f_i(t)\right) \right) dt \right] \nonumber \\
&\times \exp \left(-i \int\{ \hat{C}(t,t') x_i(t)x_i(t')   + \hat{L}(t,t') \hat{f}_i(t) \hat{f}_i(t')   + \hat{K}(t,t')  x_i(t)\hat{f}_i(t')\} dtdt' \right) \nonumber \\
&\times\left. \exp \left(-i \int \{\hat{M}(t) x_i(t)   + \hat{F}(t)i \hat{f}_i(t)  + \hat{J}(t) i \hat{f}_i(t)f_i(t) \}dt \right)Df_iD\hat{f}_i D\hat{x}_iDx_i   \right\},
\end{align}
as well as

\begin{align}
\Phi &=  \int J(t) dt
 -\mu \int F(t)M(t)dt - \frac{\sigma^2}{2}\int \{L(t,t')C(t,t') + \gamma K(t',t)K(t,t')\} dtdt',
\end{align}

and
\begin{align}
\Psi &= i \int \{\hat{C}(t,t')C(t,t') +  \hat{L}(t,t') L(t,t') + \hat{K}(t,t')K(t,t')\} dtdt'+ i \int \{ \hat{M}(t)M(t) + \hat{F}(t) F(t) + \hat{J}(t)J(t)\}dt.
\end{align}
The expression in $\Omega$ describes the time evolution of paths; this will be discussed further below. The term $\Phi$ results from the disorder average, and $\Psi$ originates from the introduction of macroscopic order parameters.

\subsection{Saddle-point integration}
In the limit of $N\to\infty$, we can evaluate the integral in Eq.~(\ref{eq:zbar}) using the saddle-point approximation. To do this, we find the values of the dynamical order parameters at which the exponent in the integral becomes extremal. 

Carrying out this extremisation, we find

\begin{align}
\pdv{\Phi}{C(t,t')} + \pdv{\Psi}{C(t,t')} = 0
\quad &\implies \quad
i\hat{C}(t,t') = \frac{\sigma^2}{2} L(t,t'), \label{hat1} \\[1em]
\pdv{\Phi}{K(t,t')} + \pdv{\Psi}{K(t,t')} = 0
\quad &\implies \quad
i\hat{K}(t,t') = \gamma \sigma^2 K(t',t),\label{hat2} \\[1em]
\pdv{\Phi}{L(t,t')} + \pdv{\Psi}{L(t,t')} = 0
\quad &\implies \quad
i\hat{L}(t,t') = \frac{\sigma^2}{2} C(t,t'),\label{hat3} \\[1em]
\pdv{\Phi}{M(t)} + \pdv{\Psi}{M(t)} = 0 \quad &\implies \quad i\hat{M}(t) = \mu F(t),\label{hat4} \\[1em]
\pdv{\Phi}{F(t)} + \pdv{\Psi}{F(t)} = 0 \quad &\implies \quad i\hat{F}(t) = \mu M(t),\label{hat5} \\[1em]
\pdv{\Phi}{J(t)} + \pdv{\Psi}{J(t)} = 0 \quad &\implies \quad i\hat{J}(t) = -1.\label{hat6}
\end{align}

\begin{align}
\pdv{\Omega}{\hat{C}(t,t')} + \pdv{\Psi}{\hat{C}(t,t')} = 0 \quad &\implies \quad C(t,t') = \lim_{N\to\infty} \frac{1}{N} \sum_i \left< x_i(t)x_i(t') \right>_\Omega,\label{sad1}\\[1em]
\pdv{\Omega}{\hat{K}(t,t')} + \pdv{\Psi}{\hat{K}(t,t')} = 0 \quad &\implies  \quad K(t,t') = \lim_{N\to\infty} \frac{1}{N} \sum_i \left< x_i(t)\hat{f}_i(t') \right>_\Omega,\label{sad2}\\[1em]
\pdv{\Omega}{\hat{L}(t,t')} + \pdv{\Psi}{\hat{L}(t,t')} = 0 \quad &\implies  \quad L(t,t') = \lim_{N\to\infty} \frac{1}{N} \sum_i \left< \hat{f}_i(t)\hat{f}_i(t') \right>_\Omega,\label{sad3}\\[1em]
\pdv{\Omega}{\hat{M}(t)} + \pdv{\Psi}{\hat{M}(t)} = 0 \quad &\implies \quad M(t) = \lim_{N\to\infty}\frac{1}{N} \sum_i \left<x_i(t)\right>_\Omega,\label{sad4}\\[1em]
\pdv{\Omega}{\hat{F}(t)} + \pdv{\Psi}{\hat{F}(t)} = 0 \quad &\implies \quad F(t) = \lim_{N\to\infty} \frac{i}{N} \sum_i \left< \hat{f}_i(t) \right>_\Omega,\label{sad5}\\[1em]
\pdv{\Omega}{\hat{J}(t)} + \pdv{\Psi}{\hat{J}(t)} = 0 \quad &\implies \quad J(t) = \lim_{N\to\infty}\frac{i}{N} \sum_i \left< \hat{f}_i(t) f_i(t) \right>_\Omega.\label{sad6}
\end{align}
In these expressions we have introduced the notation 
\begin{align}\label{avomega}
\left<A\right>_\Omega =  \frac{\int p(x(0)) A \exp\left(\omega\right)DfD\hat{f}D\hat{x}Dx}{ \int p(x(0)) \exp \left(\omega\right)DfD\hat{f} D\hat{x}Dx} ,
\end{align}
where
\begin{align}
\omega &= i \int \psi(t)x(t) dt  + i \int \hat{x}(t)  \left( \frac{\dot{x}(t)}{x(t)} -1 +x(t) - g\left(f(t)\right)\right) dt  -i \int \hat{f}(t) h(t) dt\nonumber \\
&- i \int \{\hat{C}(t,t') x(t)x(t')   + \hat{L}(t,t') \hat{f}(t) \hat{f}(t')   + \hat{K}(t,t')  x(t)\hat{f}(t')\} dtdt'\nonumber  \\
&- i \int \{\hat{M}(t) x(t)   + \hat{F}(t)i \hat{f}(t)  + \hat{J}(t) i \hat{f}(t)f(t)\} dt
\end{align}
is the argument of the exponential in $\Omega$ [Eq.~(\ref{eq:omega})],
\begin{align}
\Omega &= \ln \left[ \int p(x(0)) \exp \left(\omega\right)DfD\hat{f} D\hat{x}Dx  \right].
\end{align}
We have assumed that $h_i(t) = h(t)$ and $p(x_i(0)) = p(x(0))$ $\forall i$, we find that the terms in the sum in Eq.~(\ref{eq:omega}) are all identical, allowing us to drop the subscripts.

\subsection{Further simplification}
By differentiating this expression for the generating functional, taking the limit of $N \to \infty$, and comparing with Eqs.~(\ref{diff1})-(\ref{diff5}), we get 
\be
\overline{\left<x_i(t)\right>} = -i\left. \pdv{\overline{Z[\pmb{\psi}]}}{\psi_i(t)}\right|_{\pmb{\psi} = \mathbf{0}} = \left<x_i(t)\right>_{\Omega[\pmb{\psi} = \mathbf{0}]},
\ee
and hence,
\be
M(t) = -i \left.\lim_{N \to \infty} \frac{1}{N}\sum_i \pdv{\overline{Z[\pmb{\psi}]}}{\psi_i(t)}\right|_{\pmb{\psi} = \mathbf{0}}.
\ee
Similarly, we have

\be\overline{\left<x_i(t)x_i(t')\right>} = \left.-\pdv{\overline{Z[\pmb{\psi}]}}{\psi_i(t)}{\psi_i(t')}\right|_{\pmb{\psi} = \mathbf{0}} = \left<x_i(t)x_i(t')\right>_{\Omega[\pmb{\psi} = \mathbf{0}]},
\ee
and from this
\be
C(t,t') = - \left.\lim_{N \to \infty} \frac{1}{N}\sum_i \pdv{\overline{Z[\pmb{\psi}]}}{\psi_i(t)}{\psi_i(t')}\right|_{\pmb{\psi} = \mathbf{0}}.
\ee
Next, we use
\be
\pdv{\overline{\left<x_i(t)\right>}}{h_i(t')} = \left.-i\pdv{\overline{Z[\pmb{\psi}]}}{\psi_i(t)}{h_i(t')}\right|_{\pmb{\psi} = \mathbf{0}} = -i \left<x_i(t)\hat{f}_i(t')\right>_{\Omega[\pmb{\psi} = \mathbf{0}]},
\ee 
and have
\be
K(t,t') = \lim_{N\to\infty} \frac{1}{N}\left. \sum_i \pdv{\overline{Z[\pmb{\psi}]}}{\psi_i(t)}{h_i(t')}\right|_{\pmb{\psi} = \mathbf{0}}.
\ee
Finally, we have
\begin{align}
\hspace{-2cm}\pdv{\overline{Z[\pmb{\psi} = \mathbf{0}]}}{h_i(t)} = -i\left<\hat{f}_i(t)\right>_{\Omega[\pmb{\psi} = \mathbf{0}]} = 0 \quad &\implies \quad
F(t) = 0, \\[1em]
\hspace{-2cm}\pdv{\overline{Z[\pmb{\psi} = \mathbf{0}]}}{h_i(t)}{h_i(t')} = -\left<\hat{f}_i(t)\hat{f}_i(t')\right>_{\Omega[\pmb{\psi} = \mathbf{0}]} = 0 \quad &\implies \quad
L(t,t') = 0.
\end{align}
These equations demonstrate that taking the average over the random matrix elements in the limit $N\to\infty$ gives the same result as taking the average with respect to $\Omega$ as described above in Eq.~(\ref{avomega}), and setting $\pmb{\psi} = \mathbf{0}$. This average gives a generating functional for a dynamics of an effective process for a single species; we will discuss this next.
\subsection{Effective single-species process}
 We insert our results from the saddle-point method, Eqs.~(\ref{hat1}) to (\ref{hat6}), and find
\begin{align}
\left<A[x]\right>_{\Omega[\psi=0]} = \frac{\int A[x]P[x] Dx}{\int P[x]Dx},
n\end{align}
where
\begin{align}
P[x] &= p(x(0)) \exp\left(-i\int\hat{f}(t)h(t)dt\right)\exp\left[i\int\hat{x}(t)\left(\frac{\dot{x}(t)}{x(t)} -1 + x(t) - g(f(t))\right)dt\right] \nonumber\\
&\times \exp\left(-\int \left\{\frac{\sigma^2}{2}C(t,t')\hat{f}(t)\hat{f}(t') + \gamma\sigma^2K(t',t)x(t)\hat{f}(t')\right\}dtdt'\right) \nonumber\\
&\times \exp\left(-i\int\{\mu M(t)\hat{f}(t) - \hat{f}(t)f(t)\}dt\right) DfD\hat{f}D\hat{x}.
\end{align}
This is the probability of observing a path of the effective process given by the evolution equation
\begin{align}
\dot{x}(t) = x(t)\left[1-x(t) + g\left(\mu M(t) + \gamma\sigma^2 \int_0^t G(t,t')x(t')dt' + h(t) + \eta(t)\right)\right],
\end{align}
where
\begin{subequations}\label{supsc}
\begin{align}
M(t) &= \left<x(t)\right>_*, \\
\left<\eta(t)\eta(t')\right>_* &= \sigma^2C(t,t') = \sigma^2\left<x(t)x(t')\right>_*,\\
G(t,t') &= -iK(t,t') = \left<\pdv{x(t)_*}{h(t')}\right>.
\end{align}
\end{subequations}

In these expressions $\left<\dots\right>_*$ is the average over the effective process.
The statistics of realisations of this effective process are the same as those of the individual species trajectories in the original model. Hence we can use the effective process to find fixed points of the original system, and analyse their stability.

\subsection{Fixed-point analysis}\label{sec:supfp}
We now assume the system reaches a stable unique fixed point, $x(t) \to x^*$, which does not depend on initial conditions. This means that the macroscopic order parameters will tend to a constant, $M(t) \to M^*$ and $C(t+\tau,t) \to q$ $\forall \tau$ as $t \to \infty$. At a fixed point the response function becomes time-translation invariant, $G(t,t') = G(t-t')$, and a perturbation from a stable fixed point will have no long term effects; the integrated response function remains finite, $\chi = \int_0^\infty G(\tau) d\tau < \infty$ (see. e.g. \cite{diederich}). Each realization of $\eta(t)$ tends to a static Gaussian random variable $\eta^*$ at large times, with $\left<\eta^*\right> = 0$ and $\left<\eta^{*2}\right> = \sigma^2q$. We can then write $\eta$ as $\sigma\sqrt{q}z$ where $z$ is a standard Gaussian random number. We can now set $h(t)$ to zero, and instead generate response functions from $\eta(t)$ via
\be
G(t,t') =\left<\pdv{x(t)}{\eta(t')}\right>_*.
\ee
Fixed points satisfy the condition
\be
x^*\left[1-x^*+g\left(\mu M^* + \gamma\sigma^2\chi x^* + \sigma\sqrt{q}z\right)\right]= 0.
\ee
For any value of $z$, this relation has the solution $x^*(z) = 0$. Other solutions can be found from
\be
g\left(\mu M^* + \gamma\sigma^2\chi x^* + \sigma\sqrt{q}z\right) = x^* - 1,\label{eq:geqxminus1}
\ee
and are only valid if they are non-negative ($x^*$ represents the abundance of the effective species). 
We now evaluate this for the clipped feedback defined by
\begin{align}
g(u) = g_P(u)=
\begin{cases}
      a & u \geq a \\
      u & -a\leq u\leq a \\
      -a & u \leq -a.
 \end{cases}
 \end{align}
 The function consists of three segments, we look at each of these individually.
 
 \begin{enumerate}
\item[(i)]
If $\mu M^* + \gamma\sigma^2 \chi x^* + \sigma\sqrt{q} z \geq a$ we have, $g=a$, and hence $a = x^* -1$ from Eq.~(\ref{eq:geqxminus1}). Substituting $x^* = a+1$ into the argument of $g$ leads to the condition
\be
\mu M^* + \gamma\sigma^2 \chi(a+1) + \sigma\sqrt{q} z \geq a,
\ee
this implies
\be
z \geq \frac{(a+1)(1-\gamma\sigma^2\chi) - (\mu M^* + 1)}{ \sigma\sqrt{q}}.
\ee
 
 For later convenience we introduce the short-hand
 \be\label{sz1}
 z_2\equiv\frac{(a+1)(1-\gamma\sigma^2\chi) - (\mu M^* + 1)}{ \sigma\sqrt{q}}.
 \ee
 
\item[(ii)]\label{sec:sat-}
 For the case $\mu M^* + \gamma\sigma^2 \chi x^* + \sigma\sqrt{q} z \leq a$, we have $g = -a$ and $x^*=1-a$. As $x^*$ cannot be negative, we substitute $x^* = (1-a)\Theta(1-a)$ into the argument of $g$ to find
\be
\mu M^* + \gamma\sigma^2 \chi(1-a)\Theta(1-a) + \sigma\sqrt{q} z \leq -a
\ee
this implies
\be z \leq
\begin{cases}
\frac{-a -\mu M^*}{\sigma\sqrt{q}} & a \geq 1, \\
~ \\
\frac{(1-a)(1-\gamma\sigma^2\chi) - (1 + \mu M^*)}{\sigma\sqrt{q}} & a \leq 1.
\end{cases}\label{eq:bottom}
 \ee

\item[(iii)]
If $-a \leq (\mu M^* + \gamma\sigma^2 \chi x^* + \sigma\sqrt{q} z) \leq a$, we have $(\mu M^* + \gamma\sigma^2 \chi x^* + \sigma\sqrt{q} z) = x^* - 1$ from Eq.~(\ref{eq:geqxminus1}). This gives 
 \begin{align}
x^* = \frac{1 + \mu M^* +\sigma\sqrt{q}z}{1-\gamma\sigma^2\chi}\Theta\left(\frac{1 + \mu M^* +\sigma\sqrt{q}z}{1-\gamma\sigma^2\chi}\right),
 \end{align}
 where the Heaviside function again ensures that the solution is non-negative.
 Substituting this into $-a \leq (\mu M^* + \gamma\sigma^2 \chi x^* + \sigma\sqrt{q} z) \leq a$, we find
 \begin{align}\label{nonsat}
 -a \leq \mu M^* + \gamma\sigma^2 \chi\frac{1 + \mu M^* +\sigma\sqrt{q}z}{1-\gamma\sigma^2\chi}\Theta\left(\frac{1 + \mu M^* +\sigma\sqrt{q}z}{1-\gamma\sigma^2\chi}\right) + \sigma\sqrt{q} z \leq a.
 \end{align}
 We can take the two possible values of the Heaviside function in turn and rearrange Eq.~(\ref{nonsat}) into conditions for $z$.
 
 \begin{enumerate}
 
 \item[(a)] For a positive argument in the Heaviside function we find
\be
x^* = \frac{1 + \mu M^* +\sigma\sqrt{q}z}{1-\gamma\sigma^2\chi},
\ee
this applies when
\be
 \frac{(1-a)(1-\gamma\sigma^2\chi)\Theta(1-a) - (\mu M^* + 1)}{\sigma\sqrt{q}} \leq z \leq \frac{(1+a)(1-\gamma\sigma^2\chi) -(\mu M^*+1)}{\sigma\sqrt{q}}.
\ee

\item[(b)]For a negative argument of the Heaviside function we have  \begin{align}\label{eq:comb}
x^* = 0 \quad \text{for} \quad \frac{-a-\mu M^*}{\sigma\sqrt{q}} \leq z \leq \frac{-1 - \mu M^*}{\sigma\sqrt{q}}.
 \end{align}

 We note that for this case to occur we require the upper limit for $z$ in Eq.~(\ref{eq:comb}) to be greater than the lower limit, requiring $a \geq 1$. 
 
 We have now found two intervals for $z$ where $x^* = 0$ for $a \geq 1$: one in Eq.~(\ref{eq:bottom}) and another in Eq.~(\ref{eq:comb}). Combing these, we find
 \begin{align}
 x^* = 0 \quad \text{for} \quad z \leq \frac{-1-\mu M^*}{\sigma\sqrt{q}} \quad \text{for} \quad a \geq 1.
 \end{align}

\end{enumerate}
\end{enumerate}

 Putting these three branches for $g$ together we find
 \begin{align}
 x^* = 
 \begin{cases}
    a+1 & z \geq z_2 \\
    \frac{1 +\mu M^* + \sigma\sqrt{q}z}{1-\gamma\sigma^2\chi} & z_1 \leq z \leq z_2 \\
    (1-a)\Theta(1-a) & z \leq z_1,
 \end{cases}
 \end{align}
 where $z_2$ is as given in Eq.~(\ref{sz1}) and \begin{align}
     z_1\equiv \frac{(1-a)(1-\gamma\sigma^2\chi)\Theta(1-a) - (1 + \mu M^*)}{\sigma\sqrt{q}}.
 \end{align}

We can then simplify and solve the self-consistency relations in Eqs.~(\ref{supsc}) to find the values of these parameters for fixed points. Writing $Dz=\frac{e^{-z^2/2}}{\sqrt{2\pi}}dz$ we find
\begin{align}\label{eq:sc1}
M^*= \int_{z_2}^\infty (a + 1) Dz + \int_{z_1}^{z_2} \frac{1 +\mu M^* + \sigma\sqrt{q}z}{1-\gamma\sigma^2\chi}Dz + \int_{-\infty}^{z_1} (1-a)\Theta(1-a) Dz
\end{align}
for the mean abundance at the fixed point,
as well as
\begin{align}\label{eq:sc2}
q = \int_{z_2}^\infty (a + 1)^2 Dz+ \int_{z_1}^{z_2} \left(\frac{1 +\mu M^* + \sigma\sqrt{q}z}{1-\gamma\sigma^2\chi}\right)^2 Dz + \int_{-\infty}^{z_1} (1-a)^2\Theta(1-a) Dz
\end{align}
for the second moment of species abundances ($q=\avg{x^2}_*$). The integrated response is obtained as
\begin{align}\label{eq:sc3}
\chi &= \int_0^\infty G(\tau)d\tau \nonumber \\
&= \int_0^\infty \left<\pdv{x(t)}{\eta(t-\tau)}\right> d\tau \nonumber \\
&= \left<\pdv{x(\eta^*)}{\eta^*}\right> \nonumber \\
&= \int_{z_1}^{z_2} \frac{1}{1-\gamma\sigma^2\chi} Dz.
\end{align}

It is also useful to introduce $\phi=\int_{z_1}^{z_2} Dz$ as the probability of the abundance of the effective species not saturating [neither at at $x^* = 1+a$ nor at $x^*=(1-a)\Theta(1-a)$]. This quantity describes the fraction of species that do not saturate in the original dynamics. 

Using Eq.~(\ref{eq:sc3}) we have
\begin{align}\label{phi}
\phi = \chi(1-\gamma\sigma^2\chi).
\end{align}
\subsection{Linear stability analysis}\label{supstab}
We consider a small perturbation from the fixed point, and find the conditions for instability. The expressions for the macroscopic order parameters $\chi$, $\phi$, $M^*$ and $q$ in the previous section are based on the assumption of a unique stable fixed point; hence they are only valid for model parameters in which such a unique fixed point exists, and is stable. 
To carry out the linear stability analysis, we start from the effective process
\begin{align}
\dot{x}(t) = x(t)\left[1-x(t) + g\left(\mu M(t) + \gamma\sigma^2 \int_0^t G(t,t')x(t')dt' + \eta(t)\right)\right].
\end{align}
We follow the steps of \cite{opper, gallaasym} and write $x(t) = x^* + y(t)$ and $\eta(t) = \sigma\sqrt{q}z + v(t)$ with $\left<y(t)\right> = \left<v(t)\right> = 0$. We note that the order parameter $M$ experiences no contribution from the perturbation $y(t)$, as the ansatz assumes zero-average fluctuations about the fixed points. Self-consistency further requires $\avg{v(t)v(t')}=\sigma^2\avg{y(t)y(t')}$.

Substituting this ansatz into the effective process leads to
\begin{align}\label{eq:stabb}
\dot{y}(t) = (x^* + y(t))\left[1 - x^* - y(t) + g\left(\mu M^* + \gamma\sigma^2 \int_0^t G(t,t')(x^* + y(t'))dt' + \sigma\sqrt{q}z +v(t)\right)\right].
\end{align}
We investigate the stability of the two fixed points separately:
\begin{enumerate}
\item[(i)]
First we investigate the fixed point $x^*(z) = 0$. We then have
\begin{align}
\dot{y}(t) = y(t)\left[1 - y(t) + g\left(\mu M^* + \gamma\sigma^2 \int_0^t G(t,t')y(t')dt' + \sigma\sqrt{q}z +v(t)\right)\right],
\end{align}
which we linearise to obtain
\begin{align}
\dot{y}(t) = [1 + g(\mu M^* + \sigma\sqrt{q}z)]y(t).
\end{align}
We now distinguish several cases:
\begin{enumerate}
\item[(a)] For $z \geq (a-\mu M^*)/(\sigma\sqrt{q}$), we have $g(u) = a$, we find $\dot{y}(t) = (1+a)y(t)$. Hence $x^*=0$ is unstable. 

\item[(b)] For $z \leq (-a-\mu M^*)/(\sigma\sqrt{q})$, we have $g(u) = -a$, and we find $\dot{y}(t) = (1-a)y(t)$. The zero fixed point is stable when $a \geq 1$.

\item[(c)] For $(-a-\mu M^*)/(\sigma\sqrt{q}) \leq z \leq (a-\mu M^*)/(\sigma\sqrt{q})$ we have $g(u) = u$, we find $ \dot{y}(t) = (1 + \mu M^* + \sigma\sqrt{q}z)y(t)$. Hence, the zero fixed point is stable if $z \leq -(1+\mu M^*)/(\sigma\sqrt{q})$. Therefore we require $-a-\mu M^*\leq -1 - \mu M^*$ for stable fixed points of this type to exist, i.e., $a\geq 1$.

\end{enumerate}
We note that $x^*=0$ is stable only for the cases where it is the unique fixed point, which is when $z \leq z_1$ and $a \geq 1$, these conditions were met for stability in cases (b) and (c).
\item[(ii)]\label{sec:otherfp}
Now we consider the stability of the other fixed points found in the previous section. These satisfy
\begin{align}
1 - x^* + g\left(\mu M^* + x^*\gamma\sigma^2 \int_0^t G(t,t')dt' +\sigma\sqrt{q}z \right) = 0.
\end{align}
For $z \geq z_2$ or $z \leq z_1$, the functional is saturated, $|g(u)| = a$, and a slight perturbation will not change the value of the function, $g(u + du) = g(u)$. This gives $\dot{y}(t) = -y(t)x^*$ from Eq.~(\ref{eq:stabb}) to linear order. Hence the fixed point is stable.

To determine stability of the fixed in the case $z_1 \leq z \leq z_2$, we have $g(u) = u$, we follow \cite{opper} and add a noise variable $\xi(t)$ with $\left<\xi(t)\right> = 0$, and which we can take to be of unit amplitude. We then have
\begin{align}
\dot{y}(t) = (x^* + y(t))\left(1 - x^* - y(t) + \mu M^* + \gamma\sigma^2 \int_0^t G(t,t')(x^* + y(t'))dt' + \sigma\sqrt{q}z + v(t) + \xi(t)\right).
\end{align}
Linearizing, we obtain
\begin{align}
\dot{y}(t) = x^*\left(-y(t) + \gamma\sigma^2 \int_0^t G(t,t')y(t')dt' + v(t) + \xi(t)\right).
\end{align}
Carrying out a Fourier transform, we find

\begin{align}
\frac{i\omega\tilde{y}(\omega)}{x^*} = (\gamma\sigma^2\tilde{G}(\omega) - 1)\tilde{y}(\omega) + \tilde{v}(\omega) + \tilde{\xi}(\omega),
\end{align}
which rearranges to
\begin{align}
 \tilde{y}(\omega) =\frac{\tilde{v}(\omega) + \tilde{\xi}(\omega)}{\frac{i\omega}{x^*} + (1 - \gamma\sigma^2\tilde{G}(\omega))}.
\end{align}
We focus on the case $\omega = 0$ (a similar procedure is used in \cite{opper}, see also \cite{sollich, felixroy} for further discussion), noting that the integrated response is $\chi=\tilde{G}(0) = \int_0^\infty G(\tau) d\tau$. We find
\begin{align}
\left<\tilde{y}(0)\tilde{y}^*(0)\right>_{ns} = \frac{\left<\tilde{v}(0)\tilde{v}^*(0)\right>_{ns} + \left<\tilde{\xi}(0)\tilde{\xi}^*(0)\right>_{ns}}{(1 - \gamma\sigma^2\chi)^2},
\end{align}
where the subscript `ns' indicates that the average is over species with with non-saturated feedback only. As introduced above, these make up a fraction $\phi$ of all species at the fixed points. The remaining saturated species are stable and do not contribute to the the statistics of perturbations.

Noting that $\left<v(t)v(t')\right> = \sigma^2\left<y(t)y(t')\right>$, and hence $ \left<\tilde{v}(0)\tilde{v}^*(0)\right> = \sigma^2 \left<\tilde{y}(0)\tilde{y}^*(0)\right>$, we find
\begin{align}
\left<|\tilde{y}(0)|^2\right> = \frac{\phi(\sigma^2\left<|\tilde{y}(0)|^2\right> + 1)}{(1-\gamma\sigma^2\chi)^2}.\label{eq:instab1}
\end{align}
The factor of $\phi$ accounts for the fact that the above analysis only applies to species for which the feedback is not saturated.

Re-arranging Eq.~(\ref{eq:instab1}) we have
\begin{align}
\left<|\tilde{y}(0)|^2\right>  = \frac{\phi}{(1-\gamma\sigma^2\chi)^2 - \phi\sigma^2}.
\end{align}
In order for the fixed points to be stable this quantity needs to be finite. By construction, it also must be non-negative. The stable phase is therefore characterised by $(1-\gamma\sigma^2\chi)^2 > \phi\sigma^2$, and the onset of instability occurs when $(1-\gamma\sigma^2\chi)^2 = \phi\sigma^2$. Using Eq.~(\ref{phi}) we find the following critical heterogeneity of the coupling matrix elements,
\begin{align}\label{eq:instab3}
\sigma_c^2 = \frac{1}{\chi(1+\gamma)}.
\end{align}
The susceptibility $\chi$ is to be obtained from the self-consistency equations (\ref{eq:sc1}, \ref{eq:sc2}, \ref{eq:sc3}).
\end{enumerate}

\subsection{Minimum value for $\sigma_c$ in Fig.~\ref{mcrit}}\label{minmu}

The minimum value of $\sigma_c$ with respect to $\mu$ found in Fig.~\ref{mcrit} corresponds to a maximum value of the proportion of unsaturated species $\phi$. To show this, we begin with the instability condition in Eq.~(\ref{eq:instab3}), and differentiate with respect to $\mu$ (at fixed $\gamma$),
\begin{align}
2\chi\sigma_c\pdv{\sigma_c}{\mu} + \sigma_c^2\pdv{\chi}{\mu} = 0,
\end{align}
which results in $\pdv{\chi}{\mu} = 0$ for $\pdv{\sigma_c}{\mu} = 0$ and nonzero $\sigma_c$. By differentiating Eq.~(\ref{eq:chi}) with respect to $\mu$ we find that $\pdv{\chi}{\mu} = 0$ implies $\pdv{\phi}{\mu} = 0$.

\section{Numerical integration of the generalised Lotka-Volterra equations}

\subsection{Interaction matrix and integration}
For any pair $i<j$, the elements $\alpha_{ij}$ and $\alpha_{ji}$ in the interaction matrix are correlated (for $\gamma\neq 0$). To generate such matrices we proceeed as follows: for each pair $i<j$ draw two independent Gaussian random numbers $r_1, r_2 \sim N(0, 1)$ and set $\alpha_{ij} = \frac{\mu}{N} + \frac{\sigma }{\sqrt{N}}r_1$ and $\alpha_{ji} = \frac{\mu}{N} + \frac{\sigma}{\sqrt{N}}(\gamma r_1 + \sqrt{1-\gamma^2}r_2)$.  This ensures the entries have the required mean $\mu/N$, variance $\sigma^2/N$, and co-variance $\gamma\sigma^2/N$. We set the diagonal elements of the interaction matrix to zero, $\alpha_{ii}=0$. 

The generalised Lotka-Volterra system is a system of $N$ coupled ordinary differential equations. We typically choose $N$ to be in the range $100$-$300$. The system of ordinary differential equations is then integrated using a Runge--Kutta (RK4)  integration scheme, with a time step $\Delta t = 0.001$.

\subsection{Order parameters from numerical integration of the generalised Lotka--Volterra equations}

The data shown in Figs.~\ref{five2}, \ref{five2dh}, \ref{five3dh}, \ref{five3} is from numerical integration of the system with $N=200$. We initialise the system by drawing each $x_i(0)$ randomly from a uniform distribution on $[0,1]$. The integration is then carried out up to a final time $T=200$.

For each realisation, the macroscopic order parameters were calculated as $M = \frac{1}{N}\sum_i\left<x_i(t)\right>_{T}$, $q = \frac{1}{N}\sum_i\left<x_i^2(t)\right>_{T}$, where $\left<\cdots\right>_T$ denotes a time average in the stationary state; numerically this is carried out as the average over the last $1\%$ of the trajectory. This was then subsequently averaged over $20$  realisations of the random coupling matrix. 

In order to characterise the dynamic behaviour of the system, we also calculate the variance of the trajectory
\begin{align}
h = \frac{\left<\left<x_i(t)^2\right>_{T} - \left<x_i(t)\right>_{T}^2\right>_N}{\left<\left<x_i(t)\right>_{T}^2\right>_N}.
\end{align}
A similar quantity was used in \cite{gallaasym} to characterise the behaviour of random replicator systems. We use the notation $\avg{x_i}_N=N^{-1}\sum_i x_i$ to indicate averages over species in the numerical integration. In order to detect possible dependence on initial conditions (and convergence to multiple different fixed points from different initial coniditions), we have also carried out numerical work in which we draw one fixed instance of the interaction matrix, and then start two copies of this system. Labelling the coordinates of the two copies as $x_i$ and $x_i'$ respectively, we then calculate
\begin{align}
d = \frac{\left<\left< (x_i(t) - x'_i(t))^2\right>_N\right>_T}{\left<\left<x_i(t)\right>_N^2\right>_T}.
\end{align}
A non-zero value of this quantity at long times indicates that dependence on initial conditions.

Broadly speaking the case $h=0, d=0$ indicates that the system converges to a fixed point and that this fixed point is unique, in the sense that it does not depend on initial conditions. The case $h=0, d>0$ indicates that system converges, but that fixed point depends on initial condition (i.e., multiple stable fixed points exist). For $h>0$ finally, the system does not converge to fixed points.

Based on measures $d$ and $h$ we can only broadly classify the behaviour of the model. For example, we have not attempted to systematically characterise potential chaotic behaviour, or to distinguish it from heteroclinic cycles or limit-cycle dynamics. To do this we would need to measure the Liapunov spectrum of the system, or at the very least the leading Liapunov exponent. This goes beyond the present work, which is focused mosly on the characterisation of the phase with unique stable fixed points, and the lines in the phase diagram where global stability breaks down. We re-iterate that the behaviour shown in the phase diagrams in the main paper indicates the typical outcome. Different realisations may well show different outcomes, and it is also possible that trajectories reach a stable fixed point after a long, volatile transient.

\subsection{Species abundance distributions}
\begin{figure}[t!]
    \centering
    \includegraphics[width=0.85\textwidth]{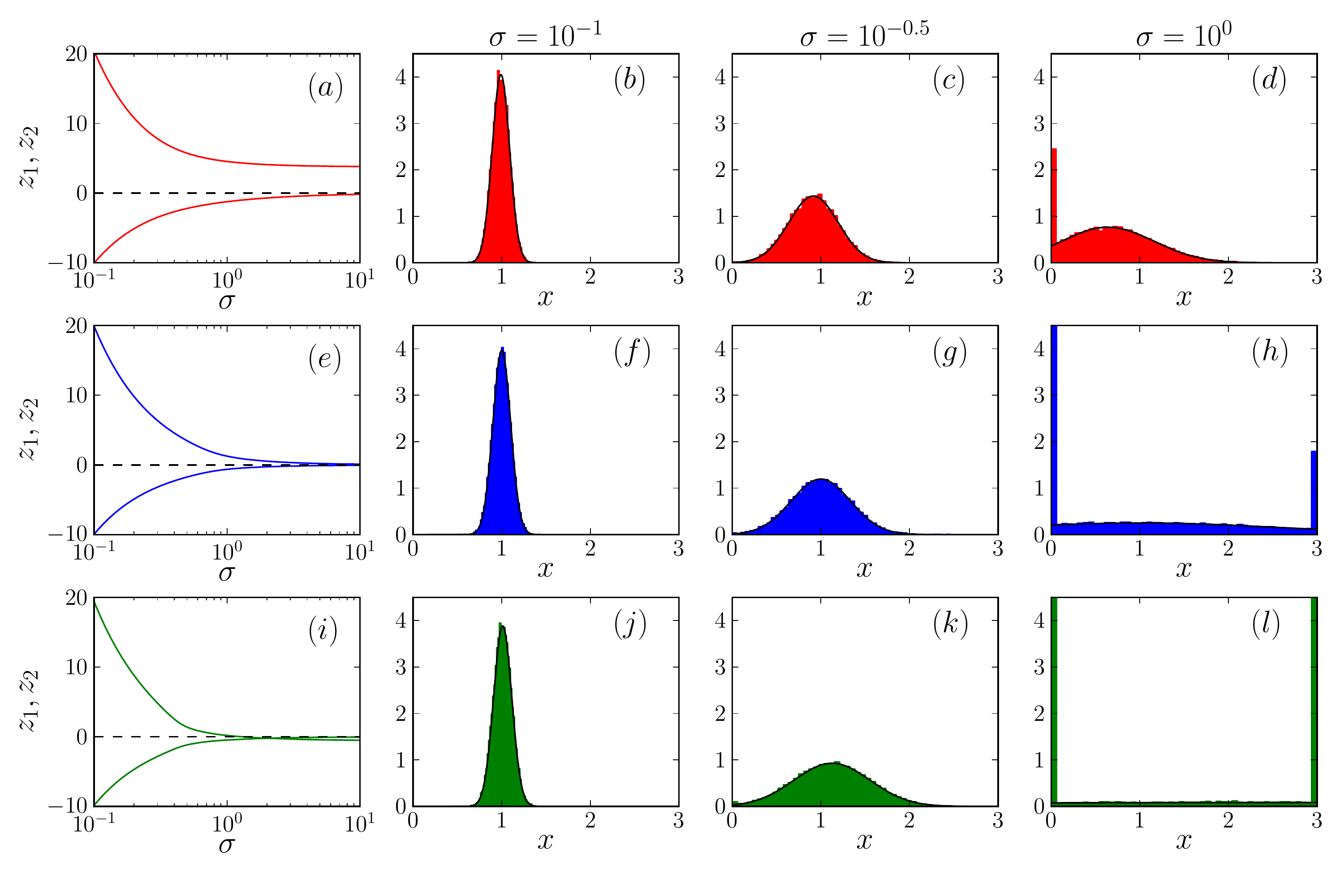}
    \caption{Species abundance distribution for the case $a=2$ and $\mu=0$ for different values of the heterogeneity $\sigma$. The upper row [panels (a)-(d)] is for $\gamma=-1$, the middle row [(e)-(h)] for $\gamma=0$, and the lower row [(i)-(l)] for $\gamma=1$. On the left [(a),(e),(i)] we show $z_2$ (upper line) and $z_1$ (lower line), remaining panels show the species abundance distributions, for $\sigma=10^{-1}$ [(b),(f),(j)], $\sigma=10^{-0.5}$ [(c),(g), (k)] and $\sigma=1$ [(d),(h),(l)]. Solid lines are theoretical predictions for the species abundance distribution, shaded histograms are from simulations. }
    \label{hist2}
\end{figure}

In Fig.~\ref{hist2} we present histograms similar to that in Fig.~\ref{hist5} in the main paper, but for a saturation of the functional feedback at $a=2$ instead of $a=0.5$. For this case, $a$ is above the carrying capacity; species then die out before they can reach the lower saturation point of $1-a$. For this case we find the upper boundary ($1+a$) to be higher, and therefore a larger heterogeneity $\sigma$ is required to spread the distribution enough for some species to reach upper saturation.

Each of the histograms in Figs.~\ref{hist5} and \ref{hist2} was generated from $100$ simulations with $N = 200$ run up to final time $T=200$.

\subsection{Phase diagrams}

To plot the phase diagrams in the main paper (Figs.~\ref{holheat} and \ref{pheat}), we simulated the system for $N=200$ for each value of $\gamma$ and $\sigma$, up to final time $T=200$. If, at any time during the simulation, any of the species abundances became greater then $10^5$, then simulation was stopped and the system was classified as divergent. If after the simulation had completed, no species abundance $x_i$ changed by more than $0.0001$ in any integration step during the last $1\%$ of the run, $|x_i(t) - x_i(t-1)| < 0.0001 \quad \forall \quad 1 \leq i \leq N, \quad 0.99T \leq t \leq T$, then the system was classified as fixed. If this condition was not met then we carried out the following additional check for convergence: if the change in abundance for each species was constantly decreasing during the last $1\%$ of the run, $|x_i(t) - x_i(t-1)| < |x_i(t-1) - x_i(t-2)|$ for all $1 \leq i \leq N$ and $0.99T \leq t \leq T$, then we assumed that it would converge to a fixed point, so was also classed as fixed.

If the system was found to converge to a fixed point then it was checked for uniqueness; this was done as follows. For each realisation of the interaction matrix, two trajectories were simulated with random initial conditions, $x_i(0), x_i'(0)$, drawn from the interval $[0, 1]$ with uniform distribution.
If $|x_i(t) - x_i'(t)| < 0.0001$ for all $1 \leq i \leq N$ and $0.99T \leq t \leq T$ then the fixed point was classified as unique. If this condition was not met then we carried out the following additional check: if the difference between each species abundance was constantly decreasing during the last $1\%$ of the run,  $|x_i(t) - x_i'(t)| < |x_i(t-1) - x_i'(t-1)|$ for all $1 \leq i \leq N$ and all $0.99T \leq t \leq T$, then we assumed that they will converge to the same value, so the fixed point was classified as unique. 

For each combination of model parameters $200$ realisations of the interaction matrix were simulated, and had its behaviour classified. The behaviours observed were plotted as a RGB colour code: the red component represents the proportion of realizations that converged to a unique fixed point, the green component that of runs with multiple fixed points, and the blue component non convergent dynamics.

\section{Numerical solution of self-consistent equations for macroscopic order parameters}

\subsection{Finding fixed point values for given $z_1$ and $z_2$}

The self consistency equations (\ref{eq:opfp}) cannot be solved directly for a given choice of model parameters $\sigma, \gamma, a$ and $\mu$. This is because one would first need to evaluate the quantities $z_1$ and $z_2$, which are present on the right-hand side of Eqs.~(\ref{eq:opfp}).  At the same time, $z_1$ and $z_2$  are functions of the order parameters we wish to find, see Eqs.~(\ref{eq:zz}).

We therefore take a parametric approach. We can solve Eqs.~(\ref{eq:opfp}) for the set of variables $\{M^*, \chi, q, \phi, \gamma, \sigma\}$ for given values of $\{z_1, z_2, a, \mu\}$. Using Eqs.~(\ref{eq:zz}), we first introduce new parameters $p_1$ and $p_2$,
\begin{align}
p_1 &\equiv \frac{(1-\gamma\sigma^2\chi)}{\sigma\sqrt{q}} = \frac{z_2 - z_1}{1 + a - (1-a)\Theta(1-a)},\nonumber \\
p_2 &\equiv \frac{1+\mu M^*}{\sigma\sqrt{q}} = \frac{(1-a)\Theta(1-a)z_2 - (1+a)z_1}{1 + a - (1-a)\Theta(1-a)}.\label{eq:pp}
\end{align}
These can be found for given $z_1$ and $z_2$. Next, we can express the self-consistency relations in Eqs.~(\ref{eq:opfp}) in terms of $p_1$ and $p_2$, to find $M^*$, $q$ and $\phi$. We have
\begin{align}
M^* &= \frac{1+a}{2}\left[1 - \erf\left(\frac{z_2}{\sqrt{2}}\right)\right] + \frac{p_2}{2p_1} \left[\erf\left(\frac{z_2}{\sqrt{q}}\right) - \erf\left(\frac{z_1}{\sqrt{2}}\right)\right] \nonumber\\
&+ \frac{1}{\sqrt{2\pi}p_1}\left[\exp\left(\frac{-z_1^2}{2}\right) - \exp\left(\frac{-z_2^2}{2}\right)\right] + \frac{(1-a)\Theta(1-a)}{2}\left[1 + \erf\left(\frac{z_1}{\sqrt{2}}\right)\right],
\end{align}
as well as
\begin{align}
q &= \frac{(1+a)^2}{2}\left[1 - \erf\left(\frac{z_2}{\sqrt{2}}\right)\right] + \frac{p_2^2 + 1}{2p_1^2}\left[\erf\left(\frac{z_2}{\sqrt{2}}\right) - \erf\left(\frac{z_1}{\sqrt{2}}\right)\right] \nonumber \\
&+ \frac{p_2}{\sqrt{2\pi}p_1^2}\left[\exp\left(\frac{-z_1^2}{2}\right) - \exp\left(\frac{-z_2^2}{2}\right)\right] - \frac{1+a}{\sqrt{2\pi}p_1}\exp\left(\frac{-z_2^2}{2}\right)\nonumber \\
&+ \frac{(1-a)\Theta(1-a)}{\sqrt{2\pi}p_1}\exp\left(\frac{-z_1^2}{2}\right) + \frac{(1-a)^2\Theta(1-a)}{2}\left[1 - \erf\left(\frac{z_1}{\sqrt{2}}\right)\right],
\end{align}
and
\begin{align}
\phi = \frac{1}{2}\left[\erf\left(\frac{z_2}{\sqrt{2}}\right) - \erf\left(\frac{z_1}{\sqrt{2}}\right)\right].
\end{align}
Now that we have $M^*$ and $q$, we can find $\sigma$ via
\begin{align}
\sigma = \frac{1 + \mu M^*}{p_2\sqrt{q}}.
\end{align}
We define another new parameter $k \equiv 1-\gamma\sigma^2\chi$ which we evaluate using $\sigma$ via
\begin{align}
k \equiv 1-\gamma\sigma^2\chi = p_3\sigma\sqrt{q},
\end{align}
which can now be used to find $\chi$ and $\gamma$ via
\begin{align}
\chi = \frac{\phi}{2k},
\end{align}
\begin{align}
\gamma = \frac{1-k}{\chi\sigma^2}.
\end{align}

In Fig.~\ref{five2} we plot graphs for the quantities $\phi$, $M$, as well as the diversity for varying values of $\sigma$ while keeping the symmetry parameter $\gamma$ fixed. The predictions for from the theory were extracted from Eqs.~(\ref{eq:opfp}) as follows:
Let the function $g(z_1, z_2, a, \mu)$ denote the value for $\gamma$ that we obtain from the algorithm described in the previous section. For any particular value of $z_2$ there should be a unique value of $z_1$ such that $g(z_1, z_2, a, \mu) = \gamma_f$, where $\gamma_f$ is the desired fixed value of the symmetry parameter. We can find this value for $z_1$ by finding the root of the function $G(z_1) = g(z_1, z_2, a, \mu) - \gamma_f$ using the Newton--Raphson method.
To plot the lines in Figs.~\ref{five2} and \ref{five3}, we chose values of $\gamma_f, a, \mu$, and varied $z_2$ to get a values of $\phi$, $M$ and diversity for a range of $\sigma$. The solutions were then parametrically plotted as a function of $\sigma$.

\subsection{Numerical identification of the instability line for continuous $a$}

To find the critical value of the interaction strenth, $\sigma_c$, for fixed parameters $\mu, a, \gamma$ shown in Fig.~\ref{acrit}, we set fixed values for the parameters $\mu$ and $a$, and found values $z_1$ and $z_2$ resulting in the desired value of $\gamma$ while satifsying Eq.~(\ref{eq:instab3}). In principle this can be done by scanning the two-dimensional plane spanned by $z_1$ and $z_2$. Instead we used a two dimensional Newton-Raphson algorithm. This consisted of a Newton-Raphson on $z_1$ to find the correct $\gamma$ inside a Newton-Raphson procedure for $z_2$ to satisfy Eq.~(\ref{eq:instab3}).

For the outer Newton-Raphson on $z_2$, let the function $s(z_2, \gamma, a, \mu)$ denote the value for $\chi\sigma^2$ that we obtain from finding the corresponding $z_1$ for $\gamma$, at given $z_2$. The critical value for $\sigma$ fulfills $\chi\sigma_c^2 = 1/(1+\gamma)$, so instability occurs when $s(z_2,\gamma,a,\mu) = 1/(1+\gamma)$. We found that  $s(z_2, \gamma, a, \mu)$ contained a singularity, so instead we took the inverse of this function, and used Newton-Raphson to find the root of $S(z_2) = 1/s(z_2, \gamma, a, \mu) - (1+\gamma)$. 

For each value of $z_2$ that $S(z_2)$ was evaluated for during the algorithm, a corresponding value for $z_1$ had to be found to obtain the required $\gamma$. The solution for $z_1$ found for a particular value of $z_2$ was used as an initial value to find $z_1$ for the next value of $z_2$ in the outer Newton-Raphson algorithm.

\subsection{Numerical identification of the instability line for different values of $\mu$}

We found that we were not able to use the algorithm described above to generate Fig.~\ref{mcrit} as the function $G(z_1) = g(z_1, z_2, a, \mu) - \gamma_f$ was not well behaved for $\mu \neq 0$. We found that sometimes the function was not smooth enough to find the root easily, or roots were found to be close to a singularity complicating the numerical evaluation of the gradient of the function $G$, required during the Newton Raphson procedure. In some instances the function $G(z_1)$ had multiple roots, with only one of them describing the physical solution. 

Instead of using the Newton-Raphson method to find a specific value for $\gamma$, we used the following method to find the original instability criteria given in Eq.~(\ref{eq:instab}): For each $\mu$ we looped through values of $z_2$ in the interval $[-10, 10]$, and for each $z_2$ we used Newton-Raphson to find the corresponding $z_1$ that satisfied the equivalent instability condition $p_1^2q - \phi = 0$. We then found the values for $\gamma$ and $\sigma$ obtained for these values of $z_1$ and $z_2$, and plotted $\sigma$ as a function of $\gamma$ for each $\mu$. To plot Fig.~\ref{mcrit} we used values of $\gamma$ to be linearly spaced with intervals of $0.01$. If the algorithm produced a value of $\gamma$ within $0.001$ of the required value, the corresponding $\sigma_c$ was accepted as a suitable approximation. We found that for values of $\mu < -4$, an extremely small change in $z_2$ produced a large change in $\gamma$. This meant that the limits of machine precision resulted inaccurate values for $\gamma$ when $\mu < -4$. The range of $\mu$ is therefore limited to $\mu \geq -4$ in Fig.~\ref{mcrit}.

%\bibliographystyle{phcpc}
%\bibliography{references}

\end{document}